\documentclass[a4paper,11pt]{article}
\usepackage{jheppub} 
\usepackage{graphicx, graphics, color}
\usepackage[T1]{fontenc} 

\usepackage{amsmath, amssymb, bm, graphicx, graphics, color, mathrsfs}
\usepackage{epsfig}
\usepackage{dcolumn}
\usepackage{bm}
\usepackage{tikz}
\usepackage{graphicx, graphics, color}
\usepackage{dcolumn}
\usepackage{bm}
\usepackage{hyperref}
\usepackage[mathlines]{lineno}

\newcommand{\be}{\begin{equation}}
\newcommand{\ee}{\end{equation}}
\newcommand{\ben}{\begin{eqnarray}}
\newcommand{\een}{\end{eqnarray}}

\newcommand{\cO}{{\cal O}}

\newcommand{\p}{\partial}
\newcommand{\na}{\nabla}

\newcommand{\Lie}{{\cal L}}

\newcommand{\tG}{\tilde G}
\newcommand{\tQ}{\tilde Q}

\newcommand{\ep}{\epsilon}

\newcommand{\ga}{\gamma}

\newcommand{\tB}{{\tilde B}}

\title{\boldmath 
Holographic calculation of the magneto-transport coefficients in Dirac semimetals}

\author[1]{Marek Rogatko\note{rogat@kft.umcs.lublin.pl, marek.rogatko@poczta.umcs.lublin.pl}}
\author[2]{Karol I. Wysokinski\note{karol@tytan.umcs.lublin.pl}}
\affiliation{Institute of Physics \\
Maria Curie-Sk{\l}odowska University \\
20-031 Lublin, pl. Marii Curie-Sk{\l}odowskiej 1, Poland}

\abstract{
Based on the gauge/gravity correspondence we have calculated  the thermoelectric 
kinetic and transport characteristics  of the strongly interacting materials 
in the presence of perpendicular magnetic field. The 3+1 dimensional system 
with Dirac-like spectrum is considered as a strongly interacting one if it is close to the  
particle-hole symmetry point. Transport in such system has been modeled by the two interacting vector 
fields. In the holographic theory  the momentum relaxation is caused by axion field and leads to finite 
values of the direct current transport coefficients. We have calculated conductivity tensor in the presence
of mutually perpendicular electric and magnetic fields and temperature gradient. The  geometry 
differs from that  in which magnetic field lies in the same plane as an electric one and temperature 
gradient. }

\keywords{Gauge-gravity correspondence,
Holography and condensed matter physics (AdS/CMT), Black Holes}

\begin{document} 

\maketitle
\flushbottom



\section{Introduction}
\label{sec:intro}
Holographic attitude \cite{mal98,wit98,gub98} in studying strongly correlated systems
offers a deep insight into their equilibrium and non-equilibrium properties \cite{zaanen-book} like 
superconductivity \cite{har08}, pseudo-gap \cite{gubankova2015}, viscosity \cite{kovtun2005} 
or thermo-electric transport. 
Recently, a great resurgence of the interests in holographic Q-lattice 
studies of the thermoelectric DC transport has been observed.
Braking  the translation invariance {by the axionic field} provides the mechanism 
of momentum dissipation in the underlying field theory and disposes to the finite values of DC
kinetic coefficients including thermoelectric matrix elements.

The number of results have already been obtained by this technique \cite{bla13}-\cite{kim15} 
for a similar model of dissipation 
and valid in principle for arbitrary value of temperature and the strength of momentum dissipation. 
The massive gravity electrical conductivity was analyzed in \cite{bla13}-\cite{dav13} 
and the consecutive  generalization to the lattice models appeared \cite{bla14}-\cite{don14a}. 
The linear axions disturbing  the translation invariance were elaborated \cite{and14}
and the thermal conductivities calculated \cite{don14b}-\cite{amo15}.
 
It was also shown that for Einstein-Maxwell scalar field gravity, 
the thermoelectric DC conductivity of the dual field theory can be achieved 
by considering a linearized Navier-Stokes equations on the black hole event horizon \cite{don15}-\cite{don16}. 
The studies in question were generalized to higher derivative gravity, which emerged 
due to the perturbative effective expansion of the string action \cite{don17}. 
The exact solution for Gauss-Bonnet-Maxwell scalar field theory for holographic 
DC thermoelectric conductivities with momentum relaxation were performed in \cite{che15}.

 The  important ingredient in the study of transport properties is a magnetic field, 
which is responsible for such phenomena as the quantum Hall, the Nernst and other effects.  
The research in this direction was conducted in \cite{bla15}-\cite{kim15}. 
Important holographic generalization of the hydrodynamic approach \cite{foster2009} 
appeared  recently. Building on the hydrodynamic idea of two independent currents 
operating in the graphene close to its particle-hole symmetry point the
authors \cite{seo2017} have used two U(1) fields and analyzed the 
charge dependence of the thermal conductance in graphene. They have got very good  
quantitative agreement  with the experimental data on the density dependence of 
the thermal conductivity in graphene. 

Graphene is one atom thick layer of graphite. The low energy spectrum of electrons is linear and described by
the relativistic Dirac like equation for mass-less Fermions. For the Fermi energy at the Dirac point 
 both electrons and holes contribute to the transport in graphene. It has been  
experimentally shown \cite{crossno2016} that for the Fermi energy coinciding with Dirac point the carriers in the 
graphene behave like a strongly interacting quantum fluid. 
 Application of the arbitrarily small electric field or
temperature gradient to graphene results in the appearance of two currents - the electron one and that 
of holes and  in the linear approximation the reaction of the material 
is characterized by kinetic  coefficients (of tensorial
character if the magnetic field is simultaneously applied) fulfilling Onsager relations.

{Dirac semi-metals (DSM) - the systems we are interested in - are the three-dimensional compounds possessing linear 
spectrum around some points in the Brillouin zone \cite{young2012}, at the Fermi energy. 
They constitute three-dimensional analogous of graphene \cite{castroneto2009}. Their crystalline 
symmetry protects the nodes in the spectrum against gap formation.
The nodes are restrained from hybridization by the combination of point group, inversion  
or time reversal symmetry \cite{fan12,yang2014}.
These materials display a host of novel properties \cite{armitage2017}.  One important difference 
between the three-dimensional DSMs and two-dimensional graphene is related to the fact that 
the number of charge carriers in graphene
can be relatively easily changed by the gate voltage. In DSM this is impossible, thus 
most of the experiments with them have been performed for constant charge density 
(close to zero) as a function of magnetic field or temperature.
The transport properties of this novel class of solids with relativistic spectrum are of great interest due to their  
responses to applied electric and magnetic fields, as well as, temperature gradients.}

{Motivated by the theoretical arguments \cite{foster2009} and experimental data \cite{levitov2016} 
that electrons in graphene, close to the Dirac point, are strongly interacting quasi-particles. 
we assume that strong interaction scenario is also realized in these three-dimensional analogs of graphene. 
Due to particle-hole symmetry and the linearity of the spectrum the role of electron-electron interaction 
is severely magnified \cite{armitage2017} in both graphene and DSMs. The arguments are related to phase space constraints
and thus are valid for both families of materials. Also in three dimensional systems  
the back-scattering in a zero magnetic field is  strongly suppressed. 
This ensues  high mobility \cite{liang2015,liang2017,he2016} of carriers.
The above features authorize holographic examination of the interaction limited transport phenomena in DSM.}

{The prediction \cite{murakami2007} and subsequent discovery \cite{hsie2008} of the 3d analogs of graphene
has resulted in  a great number of  
experimental \cite{neupane2014,neupane2016,wang2012,wang2013,brahlek2012,wu2013,liu2014,su2015,liu2014a} and 
theoretical \cite{lu2017,lundgren2014,aji2012,son2013} studies of various DSM. The conductivity tensor, 
Seebeck and Nernst effects have been measured as a function of an external magnetic field. 
Both Seebeck and Nernst coefficients give additional information
on the spectrum and properties of the materials compared to the longitudinal and Hall conductivities.
It turns out that from the experimental point of view, due to the system stability, 
Cd$_3$As$_2$ single crystals \cite{liang2015} are the most studied ones (for the recent review see  \cite{he2016}).}

{In this paper, assuming interaction dominated transport 
in DSM close to the particle-hole symmetry point, we exploit the holographic approach 
to study the thermoelectric transport in the presence of magnetic field, which is perpendicular both to 
electric field and temperature gradient. 
To calculate DC-transport coefficients we generalize recent holographic papers using 
Q-lattice approach \cite{don14Q}  without \cite{don14}-\cite{don15} or with the influence of 
magnetic field  \cite{kim15}. }

{If the Fermi energy in equilibrium coincides with the Dirac point then both electrons 
and holes coexist at arbitrary small temperature and in the  equilibrium the system is particle hole symmetric. 
Application of thermodynamic forces  induces the non-equilibrium situation and   the system's reaction is
observed as the flow of charge and heat currents. Due to the presence of both electrons and holes,
two currents will appear, which in non-equilibrium state will not cancel each other. As a result, one also expects
their mutual modification, which we shall take into account. On the gravity side, we  model this fact   
by using two different interacting $U(1)$-gauge fields representing  two currents. }

{This work  extends the previous analysis \cite{seo2017} of transport in graphene in three directions. 
We consider (i) the three dimensional analogs of graphene, (ii) allow 
the interaction between the two currents and (iii)  add external magnetic field. This enables 
us to calculate the magneto-conductance $\sigma_{xx}(B)$ and magneto-resistance $\rho_{xx}$, 
the magnetic field $B$  dependent 
Seebeck coefficient $S_{xx}$, thermal conductivity $\kappa_{xx}$ and the off-diagonal elements of these 
 transport coefficients: 
  the  Hall conductivity $\sigma_{xx}$, Hall resistivity $\rho_{xy}$, the Nernst coefficient $S_{xy}$ and off-diagonal
	component of the thermal conductivity $\kappa_{xy}$. Let us repeat that in the holographic
approach the matrix of kinetic coefficients fulfills Onsager symmetry relations as it should in any linear
theory.}

{The calculated transport coefficients are expected to describe strongly interacting carriers in 
agreement with general weak-strong coupling duality \cite{sac12} of the holographic approach.
The very good agreement between our calculations and the existing experimental data
{\it a' posteriori} supports the assumption of strongly interacting fluid existing in these
materials close to the Dirac point and shows the applicability of AdS/CMT correspondence 
to study real materials.}

The paper is organized as follows. In section \ref{sec:model} we describe the 
assumed holographic action with two interacting fields which are responsible 
for the above mentioned two currents. The holographic expressions for the heat 
and charge current are discussed in sections \ref{heat-curr} and \ref{charge-currs}, respectively.
The details of the gravity content and in particular the property of
the black hole, we have to introduce in order to equip the theory with temperature, are 
also mentioned there. The relevant kinetic and transport coefficients are calculated in Section \ref{sec;kin-tr}
and the comparison of our results with the existing experimental data presented in Section \ref{sec:exp}.
We end up with summary and conclusions. 

\section{Holographic model}
\label{sec:model}
It was shown that the proper holographic description of DC conductivities 
are provided by the so called holographic Q-lattices \cite{don14,don14a},
 i.e., the stationary black hole space-time with time-like Killing vector field. From the 
mathematical theory of black hole point of view, the black hole event horizon is the so-called Killing 
horizon in the sense that the Killing vector field in question is orthogonal 
to it. The Killing horizon has been deformed by operators that brake the translation invariance of the dual CFT. 
The breaking of the translation symmetry is achieved 
by demanding that the adequate boundary conditions are imposed on the bulk fields at the AdS-space-time boundary. 
The aim of it is to receive the finite DC-like response.
The thermoelectric conductivity in DC limit, will be found by taking into account only a linear perturbations 
of the bulk fields.

In our model the gravitational action in $(4+1)$-dimensions is taken in the form  
\be
S = \int \sqrt{-g}~ d^5 x~  \bigg( R + \frac{6}{L^2} - \frac{1}{2} \na_\mu \phi_i \na^\mu \phi^i
- \frac{1}{4} F_{\mu \nu} F^{\mu \nu} - \frac{1}{4}B_{\mu \nu} B^{\mu \nu} - \frac{\alpha}{4} F_{\mu \nu} B^{\mu \nu} \bigg),
\label{sgrav} 
\ee
where $F_{\mu \nu} = 2 \nabla_{[ \mu} A_{\nu ]}$ stands for the ordinary Maxwell field strength tensor, while
the second $U(1)$-gauge field $B_{\mu \nu}$ is given by $B_{\mu \nu} = 2 \nabla_{[ \mu} B_{\nu ]}$. $\alpha$ is a coupling constant between both gauge fields.

The equations of motion  obtained from the variation of the action $S$ with respect
to the metric, the scalar and gauge fields imply
\ben 
G_{\mu \nu} &-& g_{\mu \nu}~\frac{3}{L^2}  = T_{\mu \nu}(\phi_i) + T_{\mu \nu}(F) + T_{\mu \nu}(B) 
+ \alpha~T_{\mu \nu}(F,~B),\\ \label{ff1}
\na_{\mu}F^{\mu \nu} &+& \frac{\alpha}{2}~\na_\mu B^{\mu \nu} = 0,\\ \label{bb1}
\na_{\mu}B^{\mu \nu} &+& \frac{\alpha}{2}~\na_\mu F^{\mu \nu} = 0,\\
\na_\mu \na^\mu \phi_i &=& 0,
\een
where $G_{\mu \nu}$ is the Einstein tensor. The energy momentum tensors for the 
adequate  fields are given respectively by 
\ben
T_{\mu \nu} (\phi_i) &=& \frac{1}{2} \na_{\mu} \phi_i \na_\nu \phi_i - \frac{1}{4}~g_{\mu \nu}~\na_\delta \phi_i \na^\delta \phi_i ,\\
T_{\mu \nu}(F) &=& \frac{1}{2}~F_{\mu \delta}F_{\nu}{}^{\delta} - \frac{1}{8}~g_{\mu \nu}~F_{\alpha \beta}F^{\alpha \beta},\\
T_{\mu \nu}(B) &=& \frac{1}{2}~B_{\mu \delta}B_{\nu}{}^{\delta} - \frac{1}{8}~g_{\mu \nu}~B_{\alpha \beta}B^{\alpha \beta},\\
T_{\mu \nu}(F,~B) &=& \frac{1}{2}~F_{\mu \delta}B_{\nu}{}^{\delta} - \frac{1}{8}~g_{\mu \nu}~F_{\alpha \beta}B^{\alpha \beta}.
\een
We assume that the scalar fields depend on the three spatial coordinates we shall work with
\be
\phi_i (x_\alpha) = \beta_{i \mu} x^\mu = a_i x + b_i y + c_i z,
\ee
and the dependence will be the same for all the coordinates, i.e., $a_i =b_i = c_i = \beta$. 
The scalar field (axion) leads to the translation invariance breaking
and engenders the momentum relaxation {like scattering off impurities on the condensed matter side}. 
This fact was justified and explained in a number of papers \cite{don14, don14a}.

For the gauge fields in the considered theory we assume the {following  components}
\ben
A_\mu(r) ~dx^\mu &=& a(r) ~dt + \frac{B}{2} ~(xdy-ydx),\\
B_\mu(r)~dx^\mu  &=& b(r)~dt + \frac{B_{add}}{2} ~(xdy-ydx),
\een
where by $B$ we have denoted a background magnetic field and $B_{add}$ is the magnetic field 
of the {additional $U(1)$-gauge} field coupled to the Maxwell one.

{In the following  analysis  we  consider the  line element  }
\be
ds^2 = -f(r)dt^2 + \frac{dr^2}{f(r)} + r^2~\big(dx^2 + dy^2 + dz^2 \big).
\ee
We suppose that in the case when the five-dimensional brane  solution exists, 
its event horizon is located at $r_h$ and is subject to the relation $f(r_h)=0$.
Having in mind that the traces of the energy-momenta tensors for gauge fields are equal to zero, 
the above system of equation can be rewritten in terms of the Ricci curvature tensor $R_{\mu \nu}$ as
\be
R_{\mu \nu} + g_{\mu \nu}~\frac{2}{L^2}  = \frac{1}{2} \p_\mu \phi_i~ \p_\nu \phi^i +T_{\mu \nu}(F)  +  T_{\mu \nu}(B) 
+ \alpha~ T_{\mu \nu}(F,~B) .
\ee
The explicit forms of the Einstein equations imply
\ben
3 f(r) f'(r) + r f(r) f'(r) &=& \frac{3}{2} r f(r) \Big(
a'(r)^2 + b'^2(r) + \alpha~a'(r) b'(r) \Big),\\ \nonumber
&- & 4 r f(r) + \frac{1}{8~r} \Big( B^2 + B_{add}^2+ \alpha~B~B_{add}\Big),\\
- 2~f(r) - r~f'(r) &=& \frac{1}{4} r^2~\Big(
a'(r)^2 + b'^2(r) + \alpha~a'(r) b'(r) \Big),\\ \nonumber
&+&  \frac{1}{2} \beta^2 - 2r^2  + \frac{1}{16~r^2} \Big(B^2 + B_{add}^2+ \alpha~B~B_{add}\Big) ,\\
- 3 f'(r) - r~f''(r) &=& r~\Big( \frac{1}{2}f^2(r) - 1\Big)~\Big( a'(r)^2 + b'^2(r) + \alpha~a'(r) b'(r) \Big) \\ \nonumber
&-& 4r - \frac{f^2(r)}{8~r^3} \Big(B^2 + B_{add}^2+ \alpha~B~B_{add}\Big).
\een

\section{Heat current}
\label{heat-curr}
In this section we pay attention to the definition of the heat current and the thermoelectric conductivities. 
The key point in conducting the aforementioned calculations is to find radially independent 
quantities in the bulk which can be identified with the adequate boundary currents.
Namely, having in mind the adequate Killing vector field and the equations of motion, 
one obtains the two-form which will be equal to zero when the divergence with respect to $r$-coordinate
will be performed. 

To commence with, let us suppose that   $k_\mu = (\p/\p t)_\mu$ is a time-like Killing vector field. 
We choose asymptotically time-like Killing vector field because of the fact that one considers static 
space-time for which exists space-like hyper-surface which is orthogonal to the orbits of the isometry 
generated by the aforementioned Kiling vector field.

The general properties of the Killing vector enables us to find that
\be
\na_\mu \na^\nu k^\mu = T^\nu {}{}_\mu k^\mu - \frac{k^\nu~T}{d-2} - 2 \frac{k^\nu \Lambda}{d-2},
\label{gen}
\ee
where $T =T_\mu {}{}^\mu$ denotes the trace of the energy momentum tensor and $\Lambda$ is the cosmological constant.
In the considered case, we impose the following symmetry conditions for the fields appearing in our model
\be
\Lie_k F_{\alpha \beta} = \Lie_k B_{\alpha \beta}= \Lie_k \phi = 0,
\label{lie}
\ee
where  $\Lie$ denotes the Lie's derivative with respect to the vector field $k_\mu$.
One has also that
\be
k^\mu ~F_{\mu \nu} = \na_\nu \theta_{(F)}, \qquad k^\mu ~B_{\mu \nu} = \na_\nu \theta_{(B)},
\ee
where $\theta_{(F)}$ and  $\theta_{(B)}$ are arbitrary functions. 
Having in mind the   equations of motion (\ref{ff1}) and (\ref{bb1}), as well as, 
the exact form of the Lie derivatives (\ref{lie})
for gauge fields, one arrives at relations
\ben
k^\mu ~F_{\mu \alpha}F^{\rho \alpha} = \na_\alpha \Big( \theta_{(F)}~F^{\rho \alpha} \Big), \qquad
k^\mu ~B_{\mu \alpha}B^{\rho \alpha} = \na_\alpha \Big( \theta_{(B)}~B^{\rho \alpha} \Big),\\ 
k^\mu ~F_{\mu \ga}B^{\rho \ga} + k^\mu ~B_{\mu \alpha}F^{\rho \alpha} = \na_\delta \Big( \theta_{(F)} B^{\nu \delta} \Big) +  \na_\delta \Big( \theta_{(B)} F^{\nu \delta} \Big).
\een
Moreover, it can be checked that the following set of equations is satisfied
\ben
k^\mu~F^{\rho \nu}F_{\rho \nu} &=& 4~\na_\rho \Big( k^{[\mu}F^{\rho]}A_\nu \Big) + 2~\Lie_k A_\nu~F^{\mu \nu}, \\
k^\mu~B^{\rho \nu}B_{\rho \nu} &=& 4~\na_\rho \Big( k^{[\mu}B^{\rho]}B_\nu \Big) + 2~\Lie_k B_\nu~B^{\mu \nu}, \\
k^\mu~B^{\rho \nu}F_{\rho \nu} &=& 4~\na_\rho \Big( k^{[\mu}F^{\rho]}B_\nu \Big) + 2~\Lie_k B_\nu~F^{\mu \nu}, \\
k^\mu~F^{\rho \nu}B_{\rho \nu} &=& 4~\na_\rho \Big( k^{[\mu}B^{\rho]}A_\nu \Big) + 2~\Lie_k A_\nu~B^{\mu \nu}, 
\een
Using the relation (\ref{gen}), after some algebra, one finds that
\be
\na_\rho \tG_{\nu \rho} = - 2 \frac{\Lambda~k^\nu}{d-2},
\label{2form}
\ee
where the two-form in question implies  
\ben
\tG_{ \nu \rho} = \na^\nu k^\rho &+& \frac{1}{2} \Big( k^{[\nu}F^{\rho] \alpha}A_\alpha \Big) + \frac{1}{4} \Big[ \Big( \psi - 2 \theta_{(F)} \Big)~F^{\nu \rho} \Big] \\ \nonumber
&+& \frac{1}{2} \Big( k^{[\nu}B^{\rho] \alpha}B_\alpha \Big) + \frac{1}{4} \Big[ \Big( \chi - 2 \theta_{(B)} \Big)~B^{\nu \rho} \Big] \\ \nonumber
&+& \frac{\alpha}{4} \Big[ \Big( k^{[\nu}B^{\rho] \alpha}A_\alpha \Big) + \Big( k^{[\nu}F^{\rho] \alpha}B_\alpha \Big) \Big] \\ \nonumber
&+& \frac{\alpha}{8}\Big[ \Big( \psi - 2 \theta_{(F)} \Big)~B^{\nu \rho} \Big] + \frac{\alpha}{8} \Big[ \Big( \chi - 2 \theta_{(B)} \Big)~F^{\nu \rho} \Big].
\een
In the derivation of the relation (\ref{2form}) we have used the equations provided by
\ben
\Lie_k A_\alpha~F^{\nu \alpha} = \na_\rho \Big( \psi~F^{\nu \rho} \Big), \qquad \Lie_k B_\alpha~B^{\nu \alpha} = \na_\rho \Big( \chi~B^{\nu \rho} \Big),\\
\Lie_k A_\alpha~B^{\nu \alpha} = \na_\rho \Big( \psi~B^{\nu \rho} \Big), \qquad \Lie_k B_\alpha~F^{\nu \alpha} = \na_\rho \Big( \chi~F^{\nu \rho} \Big),
\een
where we have set
\ben
\psi &=& E_\alpha x^\alpha, \qquad \theta_{(F)} = - E_\alpha x^\alpha - a(r),\\
\chi &=& B_\alpha x^\alpha, \qquad \theta_{(B)} = - B_\alpha x^\alpha - b(r).
\een
{ The symbol $E_\alpha$ denotes the component $a$ of the Maxwell electric field while $B_\alpha$
 corresponds to 'electric' field bounded to the other gauge 
field sector}, $\alpha = x,~y$.  A close inspection of (\ref{2form}) reveals that the right-hand 
side is equal to zero if one considers the Killing vector $k^\nu$
with the index different from the one connected with time coordinate. 
One can see that the $\tG_{\nu \rho}$ tensor is antisymmetric and satisfies
\be
\p_\rho \Big( 2~\sqrt{-g}~\tG^{\nu \rho} \Big) = - 2 \frac{\Lambda~\sqrt{-g}~k^\nu}{d-2}.
\ee
In our considerations we shall use the two-form given by $2~\tG_{ \nu \rho} $, i.e.,
the heat current will be defined as $Q^i = 2~\sqrt{-g}~ \tG^{ i r} $.

\section{Charge currents}
\label{charge-currs}
In this section we shall obtain the general form of the charges of the black brane in terms of its 
event horizon data. In the dual theory
the current density is of the form $J_{(F)}^\mu =\sqrt{-g} (F^{\mu r} + \alpha/{2}~ B^{\mu r} )$ and $J_{(B)}^\mu=\sqrt{-g} (B^{ \mu r} + \alpha/{2} ~F^{\mu r} )$, where
the right-hand sides are evaluated at the spacetime boundary, when $r \rightarrow \infty$. The only non-zero component of the equations of motion for the considered gauge fields are in 
time-coordinate direction. Therefore we can write that charges of the black brane calculated at any value of the $r$-coordinate, including the case where $r=r_h$, are provided by
\ben
\tQ_{(F)} &=& \sqrt{-g}~\Bigg(F^{rt} + \frac{\alpha}{2} B^{rt} \Bigg) = Q_{(F)} + \frac{\alpha}{2} Q_{(B)},\\
\tQ_{(B)} &=& \sqrt{-g}~\Bigg(B^{rt} + \frac{\alpha}{2} F^{rt} \Bigg) = Q_{(B)} + \frac{\alpha}{2} Q_{(F)},
\een
where we have set $Q_{(F)} = r^3~a'(r), ~Q_{(B)}=  r^3~b'(r)$.

In order to find the conductivities for the background in question, one takes 
into account small perturbations around the background solution obtained from Einstein equations of motion.
The perturbations imply
\ben \label{f1}
\delta A_i &=& t~\Big( - E_i + \xi_i~a(r) \Big) + \delta a_i(r),\\
\delta B_i &=& t~\Big( - B_i + \xi_i~b(r) \Big) + \delta b_i(r),\\
\delta G_{ti} &=& t~\Big( - \xi_i~f(r) \Big) + \delta g_{ti}(r),\\
\delta G_{r i} &=& r^2~\delta g_{ri}(r),\\ \label{f5}
\delta \phi_i  &=& \delta \phi_i(r),
\een
where $t$ is time coordinate. We put $i= x,~y$, and denote the temperature gradient 
by $\xi_i = - \na_i T/T$.

The {\it electric currents}  will be associated with the radially independent components 
of the equations (\ref{ff1}) and (\ref{bb1}), which in turn can be calculated everywhere
in the bulk. Because of the form the underlying equations they will constitute the mixture 
of two $U(1)$-gauge fields. Their definitions are provided by
\be
J^i_{(F)}(r) = \sqrt{-g} ~\Big( F^{ir} + \frac{\alpha}{2} B^{ir} \Big),
\ee
which implies the following:
\ben
J^i_{(F)}(r) &=& - r~\Big[
\delta^{ij}~\delta g_{tj}~\Big( a'(r) + \frac{\alpha}{2} b'(r) \Big) + f(r)~\delta^{ij} \Big( \delta a'_j(r) + \frac{\alpha}{2}~\delta b'_j(r) \Big) \\ \nonumber
&-& \ep^{ij}~ f(r) ~\Big( \frac{B}{2}  + \frac{\alpha}{2} \frac{B_{add}}{2} \Big) ~\delta g_{rj} \Big].
\een
On the other hand, the current bounded with the other gauge field is given by
\be
J^i_{(B)}(r) = \sqrt{-g} ~\Big( B^{ir} + \frac{\alpha}{2} F^{ir} \Big).
\ee
Its exact form is subject to the relation
\ben
J^i_{(B)}(r) &=& - r~\Big[
\delta^{ij}~\delta g_{tj}~\Big( b'(r) + \frac{\alpha}{2} a'(r) \Big) + f(r)~\delta^{ij} \Big( \delta b'_j(r) + \frac{\alpha}{2}~\delta a'_j(r) \Big) \\ \nonumber
&-& \ep^{ij}~ f(r) ~\Big( \frac{B_{add}}{2}  + \frac{\alpha}{2} \frac{B}{2} \Big) ~\delta g_{rj} \Big].
\een
However, the presence of magnetization causes that one should into account the non-trivial fluxes 
connected with the non-zero components $B$ and $B_{add}$. The linearized
equations describing the continuity equation of one of the U(1) fields can be written in the form 
\ben \nonumber
0 = \p_{M} \bigg[ \sqrt{-g}~\Big( F^{iM} + \frac{\alpha}{2} B^{iM} \Big) \bigg] &=& \p_r  \bigg[ \sqrt{-g}~\Big( F^{i r} + \frac{\alpha}{2} B^{i r} \Big) \bigg] \\ \nonumber
&+& \p_t  \bigg[ \sqrt{-g}~\Big( F^{i t} + \frac{\alpha}{2} B^{i t} \Big) \bigg] ,
\label{c1}
\een
and for the other gauge field the equation of motion gives 
\ben
0 = \p_{M} \bigg[ \sqrt{-g}~\Big( B^{iM} + \frac{\alpha}{2} F^{iM} \Big) \bigg] &=& \p_r  \bigg[ \sqrt{-g}~\Big( B^{i r} + \frac{\alpha}{2} F^{i r} \Big) \bigg] \\ \nonumber
&+& \p_t  \bigg[ \sqrt{-g}~\Big( B^{i t} + \frac{\alpha}{2} F^{i t} \Big) \bigg] .
\label{c2}
\een
Because of the fact that {\it electric currents} are r-independent, we shall evaluate 
them on the black object event horizon. Integrating
the above relations we arrive at the currents at the boundary of $AdS_5$
\ben
J^i_{(F)}(\infty) &=& J^i_{(F)}(r_h) + \frac{B}{2}~\ep^{ij}~\xi_j~\Sigma_{(1)} + \frac{\alpha}{2} \frac{B_{add}}{2}~\xi_j~\Sigma_{(1)},\\
J^i_{(B)}(\infty ) &=& J^i_{(B)}(r_h) + \frac{B_{add}}{2}~\ep^{ij}~\xi_j~\Sigma_{(1)} + \frac{\alpha}{2} \frac{B}{2}~\xi_j~\Sigma_{(1)} ,
\een
where one denotes by $\Sigma_{(1)} = \int_{r_h}^{\infty} dr'~\frac{1}{r'}$. Mathematically, this integral comes from  
the equations (\ref{c1}) and (\ref{c2}), right-hand side of them, and appears due to the integration
in five-dimensional spacetime, where the volume element is proportional to $\sim \sqrt{-g} =r^3$. As one can see below 
(the end of this section) all such terms should be excluded in order to achieve the DC-conductivities.

The heat current at the linearized order implies
\be
Q^i(r) = 2~\sqrt{-g}~\tG^{ri} 
= 2~\sqrt{-g}\na^r k^i - a(r)~J^i_{(F)}(r) - b(r)~J^i_{(B)}(r),
\ee
and is subject to the relation $\p_\mu [2\sqrt{-g} \tG^{\mu \nu}] = 0$, 
in the absence of a thermal gradient. But the existence of magnetization 
currents enforces  the following equations
\ben
\p_r [ 2\sqrt{-g}\tG^{rx}] &=& -\p_t [2\sqrt{-g}\tG^{tx}] - \p_y [ 2\sqrt{-g}\tG^{yx} ] - a(r)  J^x_{(F)}(\infty)  - b(r) J^x_{(B)}(\infty) ,\\
\p_r [2 \sqrt{-g}\tG^{ry}] &=& -\p_t [2\sqrt{-g}\tG^{ty}] - \p_y [ 2\sqrt{-g}\tG^{xy} ] - a(r)  J^y_{(F)}(\infty)  - b(r) J^y_{(B)}(\infty).
\een
In order to achieve the radially independent form of the current, one ought to add additional 
terms to get rid of the aforementioned fluxes.
The considered quantity should obey $\p_i \tQ^i = 0$,  then one has to have
\ben \nonumber
\tQ^i (\infty) &=& Q^i(r_h) + \frac{B}{2}~\ep^{ij}~E_j~\Sigma_{(1)} - B~\ep^{ij}\xi_j~\Sigma_{(a)} 
+ \frac{B_{add}}{2}~\ep^{ij}~E_j~\Sigma_{(1)} - B_{add}~\ep^{ij}\xi_j~\Sigma_{(a)} \\ 
&-& \frac{\alpha}{2} \Big (B~\ep^{ij}~B_j +  B_{add}~\ep^{ij}~B_j \Big) \Sigma_{(b)} 
+ \frac{\alpha}{4} \Big( B_{add}~\ep^{ij}~E_j + B~\ep^{ij}~B_j \Big) \Sigma_{(1)},
\een
where we have denoted
\be
\Sigma_{(a)} = \int_{r_h}^{\infty} dr'~\frac{a(r')}{r'}, \qquad \Sigma_{(b)} = \int_{r_h}^{\infty} dr'~\frac{b(r')}{r'}.
\ee
We have obtained three boundary currents $J^i_{(F)}(\infty),~J^i_{(B)}(\infty)$ and $\tQ^i(\infty)$, 
which can be simplified by imposing the regularity conditions at the black brane horizon.
Namely, they imply the following:
\ben \label{r1}
\delta a_i(r) &\sim& - \frac{E_i}{4~\pi~T} \ln (r-r_h) + \dots,\\ \label{r2}
\delta b_i(r) &\sim& - \frac{B_i}{4~\pi~T} \ln (r-r_h) + \dots,\\ \label{r3}
\delta g_{ri}(r) &\sim& \frac{1}{r_h^2}~\frac{\delta g^{(h)}_{ti}}{f(r_h)} + \dots,\\ \label{r4}
\delta g_{ti}(r) &\sim& \delta g^{(h)}_{ti}+ \cO(r-r_h) + \dots,\\ \label{r5}
\delta \phi_i(r) &\sim & \phi_i(r_h) +  \cO(r-r_h) + \dots,
\een
where $T=1/4 \pi~\p_r f(r)\mid_{r=r_h}$ is the Hawking temperature of the black brane in question.

The above relations lead to the following forms of the  boundary currents 
\ben \nonumber \label{jiF}
J^i_{(F)}(\infty) &=& r_h~\Bigg( - \frac{\tQ_{(F)}}{r_h^3}~\delta^{ij}~\delta g^{(h)}_{tj}  + \frac{\delta g^{(h)}_{tj}}{r_h^2} \Big(
\ep^{ij}~\frac{B}{2} + \frac{\alpha}{2}~\ep^{ij} ~\frac{B_{add}}{2} \Big) + \delta^{ij} E_j  + \frac{\alpha}{2}~\delta^{ij} B_j  \Bigg)\\ \label{jiB}
 &+& \frac{B}{2}~\ep^{ij}~\xi_j~\Sigma_{(1)} + \frac{\alpha}{2} \frac{B_{add}}{2}~\xi_j~\Sigma_{(1)},\\ \nonumber
J^i_{(B)}(\infty) &=& 
r_h~\Bigg( - \frac{\tQ_{(B)}}{r_h^3}~\delta^{ij}~\delta g^{(h)}_{tj}  + \frac{\delta g^{(h)}_{tj}}{r_h^2} \Big(
\ep^{ij}~\frac{B_{add}}{2} + \frac{\alpha}{2}~\ep^{ij} ~\frac{B}{2} \Big) +
  \delta^{ij} B_j  + \frac{\alpha}{2}~\delta^{ij} E_j  \Bigg)\\ 
 &+& \frac{B_{add}}{2}~\ep^{ij}~\xi_j~\Sigma_{(1)} + \frac{\alpha}{2} \frac{B}{2}~\xi_j~\Sigma_{(1)},\\ \nonumber
\tQ^i(\infty) &=& - 4~\pi~T~\delta^{ij}~\delta g^{(h)}_{tj} + \frac{B}{2}~\ep^{ij}~E_j~\Sigma_{(1)} - B~\ep^{ij}~\xi_j~\Sigma_{(a)} \\ \nonumber \label{qui}
&+&\frac{B_{add}}{2} ~\ep^{ij}~B_j~\Sigma_{(1)} - B_{add}~\ep^{ij}~\xi_j~\Sigma_{(a)} 
- \frac{\alpha}{2}~B~\ep^{ij}~\xi_j~\Sigma_{(b)} - \frac{\alpha}{2}~B_{add}~\ep^{ij}~\xi_j~\Sigma_{(b)}  \\ 
&+& \frac{\alpha}{4} \Big( B_{add}~\ep^{ij}~E_j + B~\ep^{ij}~B_j \Big) ~\Sigma_{(1)}.
\een
As was mentioned in \cite{har07,kim15} the terms proportional to  $\Sigma_{(j)} B_{\zeta}/T$, where $j=1,a,b$ 
and $B_\zeta = B, B_{add}$, emerge from the contributions of magnetization
currents which stem from the two considered $U(1)$-gauge fields. They should be subtracted from 
the adequate expressions for the DC-conductivities. Thus calculating the conductivities  below 
we shall neglect all such terms.

On the other hand, the linear Einstein equations for the fluctuations given by the 
relations (\ref{f1})-(\ref{f5}) are provided by
\ben \nonumber
- \frac{1}{2} f(r)~\delta g_{tx}''(r) &=& \frac{1}{2} f(r) \bigg[ \frac{B}{2} \bigg(
- E_y + \xi_y~a(r) \bigg) \frac{1}{r^2~f(r)}  - a'(r) \Big( - \delta a'_x(r) + \frac{B}{2} \delta g_{ry}(r) \Big) \\ \nonumber
&+&
\frac{B_{add}}{2} \bigg(
- B_y + \xi_y~b(r) \bigg) \frac{1}{r^2~f(r)}  - b'(r) \Big( - \delta b'_x(r) + \frac{B_{add}}{2} \delta g_{ry}(r) \Big) \\ \nonumber
&+&
\frac{\alpha}{2} \bigg[ \frac{B_{add}}{2} \bigg(
- E_y + \xi_y~a(r) \bigg) \frac{1}{r^2~f(r)}  - a'(r) \Big( - \delta b'_x(r) + \frac{B_{add}}{2} \delta g_{ry}(r) \Big) \\ \nonumber
&+&
\frac{B}{2} \bigg(
- B_y + \xi_y~b(r) \bigg) \frac{1}{r^2~f(r)}  - b'(r) \Big( - \delta a'_x(r) + \frac{B}{2} \delta g_{ry}(r) \Big) \bigg] \\ 
& -& \frac{1}{8}\delta g_{tx}(r)\bigg(
\frac{B^2}{2~r^4} + \frac{B_{add}^2}{2~r^4} + 2~\alpha~\frac{B B_{add}}{4~r^4} +
\frac{6 \beta^2}{r^2} \bigg),
 \een
and 
\ben \nonumber
- \frac{1}{2} f(r)~\delta g_{ty}''(r) &=& \frac{1}{2} f(r) \bigg[ - \frac{B}{2} \bigg(
- E_x + \xi_x~a(r) \bigg) \frac{1}{r^2~f(r)}  + a'(r) \Big(  \delta a'_x(r) + \frac{B}{2} \delta g_{rx}(r) \Big) \\ \nonumber
&-&
\frac{B_{add}}{2} \bigg(
- B_x + \xi_x~b(r) \bigg) \frac{1}{r^2~f(r)}  + b'(r) \Big(  \delta b'_x(r) + \frac{B_{add}}{2} \delta g_{rx}(r) \Big) \\ \nonumber
&+&
\frac{\alpha}{2} \bigg[ - \frac{B_{add}}{2} \bigg(
- E_x + \xi_x~a(r) \bigg) \frac{1}{r^2~f(r)}  + a'(r) \Big( \delta b'_x(r) + \frac{B_{add}}{2} \delta g_{rx}(r) \Big) \\ \nonumber
&-&
\frac{B}{2} \bigg(
- B_x + \xi_x~b(r) \bigg) \frac{1}{r^2~f(r)}  + b'(r) \Big(  \delta a'_x(r) + \frac{B}{2} \delta g_{rx}(r) \Big) \bigg] \\ 
& -& \frac{1}{8}\delta g_{tx}(r)\bigg(
\frac{B^2}{2~r^4} + \frac{B_{add}^2}{2~r^4} + 2~\alpha~\frac{B B_{add}}{4~r^4} +
\frac{6 \beta^2}{r^2} \bigg).
\een
Consequently, using the other Einstein equations, we obtain the relations governing $\delta g_{tx}$ 
and $\delta g_{ty}$. They yield
\ben \nonumber
\delta g_{rx} &=& \frac{1}{\frac{B^2+B_{add}^2}{2~r^4} + \frac{2 \alpha (B_{add} B)}{4 r^4} + \frac{6 \beta^2}{r^2}}
~\bigg[
4~\frac{a'(r)}{f(r)}~\Big( - E_x + \xi_x a(r) \Big) + \frac{2 B}{r^2}~\delta a'_y(r) \\ \nonumber
&+& 4~\frac{b'(r)}{f(r)}~\Big( - B_x + \xi_x b(r) \Big) + \frac{2 B_{add}}{r^2}~\delta b'_y(r) 
\\ \nonumber
&+& 
2 \alpha~\Big[ \frac{a'(r)}{f(r)} ~\Big( - B_x + \xi_x b(r) \Big) + \frac{b'(r)}{f(r)} ~\Big( - E_x + \xi_x a(r) \Big) \Big]- 4 \frac{\xi_x~f'(r)}{f(r)} \\ 
&-& \delta g_{ty}~\Big( \frac{2 a'(r) + \alpha b'(r)}{r^2~f(r)} B + \frac{2 b'(r) + \alpha a'(r)}{r^2~f(r)} B_{add} \Big)\bigg],
\een
\ben \nonumber
\delta g_{ry} &=& \frac{1}{\frac{B^2+B_{add}^2}{2~r^4} + \frac{2 \alpha (B_{add} B)}{4 r^4} + \frac{6 \beta^2}{r^2}}
~\bigg[
4~\frac{a'(r)}{f(r)}~\Big( - E_y + \xi_y a(r) \Big) - \frac{2 B}{r^2}~\delta a'_x(r) \\ \nonumber
&+& 4~\frac{b'(r)}{f(r)}~\Big( - B_y + \xi_y b(r) \Big) - \frac{2 B_{add}}{r^2}~\delta b'_x(r) 
\\ \nonumber
&+& 
2 \alpha~\Big[ \frac{a'(r)}{f(r)} ~\Big( - B_y + \xi_y b(r) \Big) + \frac{b'(r)}{f(r)} ~\Big( - E_y + \xi_y a(r) \Big) \Big]- 4 \frac{\xi_x~f'(r)}{f(r)} \\ 
&+&
\delta g_{ty}~\Big( \frac{2 a'(r) + \alpha b'(r)}{r^2~f(r)} B + \frac{2 b'(r) + \alpha a'(r)}{r^2~f(r)} B_{add} \Big)\bigg],\
\een
In the next step, we shall implement the near horizon black brane expressions given by the equations (\ref{r1})-(\ref{r5}) to rewrite the above relations. Namely, we use
the definitions of the charges $Q_{(F)},~Q_{(B)}$
and the relation between   $\delta g_{rj}$ and $\delta g_{tj}$ as given by (\ref{r3}). It all leads to the following:
\ben \label{g1}
\delta g^{(h)}_{tx} &=& \frac{r_h}{\frac{B^2 +B^2_{add}}{2 r_h^2 }+ \frac{2 \alpha (B_{add} B)}{4 r_h^4} + 6 \beta^2 }~
\bigg[
- 4 \Big( Q_{(F)}(r_h) ~E_x + Q_{(B)}(r_h) ~B_x \Big) \\ \nonumber
&-& 2 B  E_y ~r_h - 2 B_{add} B_y~ r_h - 16  \pi ~T~\xi_x~r_h^3  -  2 \alpha~\Big(Q_{(F)}(r_h) ~B_x + Q_{(B)}(r_h) ~E_x \Big) \\ \nonumber
&-& \delta g^{(h)}_{ty}~\Big[
\frac{B}{r_h^2}~\Big( 2 Q_{(F)}(r_h) + \alpha Q_{(B)}(r_h) \Big)  + 
\frac{B_{add}}{r_h^2}~\Big( 2 Q_{(B)}(r_h) + \alpha Q_{(F)}(r_h) \Big) \Big] \bigg],
\een
and
\ben  \label{g2}
\delta g^{(h)}_{ty} &=& \frac{r_h}{\frac{B^2 +B^2_{add}}{2 r_h^2 }+ \frac{2 \alpha (B_{add} B)}{4 r_h^4} + 6 \beta^2 }~
\bigg[
- 4 \Big( Q_{(F)}(r_h) ~E_y + Q_{(B)}(r_h) ~B_y \Big) \\ \nonumber
&+& 2 B  E_x ~r_h + 2 B_{add} B_x~ r_h - 16  \pi ~T~\xi_y~r_h^3  -  2 \alpha~\Big(Q_{(F)}(r_h) ~B_y + Q_{(B)}(r_h) ~E_y \Big) \\ \nonumber
&+& \delta g^{(h)}_{tx}~\Big[
\frac{B}{r_h^2}~\Big( 2 Q_{(F)}(r_h) + \alpha Q_{(B)}(r_h) \Big)  + 
\frac{B_{add}}{r_h^2}~\Big( 2 Q_{(B)}(r_h) + \alpha Q_{(F)}(r_h) \Big) \Big] \bigg].
\een

The solutions of the equations (\ref{g1})-(\ref{g2}) can be written in the forms as  
\ben
\delta g^{(h)}_{tx} &=& \frac{1}{A^2 + C^2}~\Big[
- A~\Big( K_x + D_x + H_x \Big) + C~\Big( K_y + D_y + H_y \Big)\Big],\\
\delta g^{(h)}_{ty} &=& \frac{1}{A^2 + C^2}~\Big[
- A~\Big( K_y + D_y + H_y \Big) - C~\Big( K_x + D_x + H_x \Big) \Big],
\een
where we have set for $m=x,~y$, and the coefficients appearing in  the relations are defined by
\ben
\label{def-a}
A &=& \frac{\tB^2+ 12 \beta^2~r_h^2 }{2 ~r_h^3},\\
\label{def-b}
K_m &=& 4\bigg( Q_{(F)}(r_h) ~E_m + Q_{(B)}(r_h) ~B_m \bigg) + 2 \alpha \bigg( Q_{(F)}(r_h) ~B_m + Q_{(B)}(r_h) ~E_m \bigg),\\
\label{def-c}
C &=& \frac{B}{r_h^2}~\bigg( 2 Q_{(F)}(r_h) + \alpha Q_{(B)}(r_h) \bigg)
+  \frac{B_{add}}{r_h^2}~\bigg( 2 Q_{(B)}(r_h) + \alpha Q_{(F)}(r_h) \bigg),\\
\label{def-d}
D_m &=& 16 \pi ~T~\xi_m~r_h^3,\\
H_m &=& - 2~B~\ep_{m k} ~E^k~r_h  - 2~B_{add}~\ep_{m k}~ B^k~r_h.
\label{def-h}
\een
For the brevity of the notation we set $\tB^2 = B^2 + B_{add}^2 + \alpha B B_{add}$.

The explicit analysis of the model requires dyonic black hole solutions. 
The solution describing asymptotically flat and non-flat dyonic black hole 
with the topology $S^3$ of the event horizon 
were achieved by the generation technique \cite{yaz06} from dyonic black ring solution 
or from Thangherlini black hole. The obtained asymptotically flat solution was in fact 
five-dimensional Gibbons-Maeda dyonic black hole \cite{gib88} attained by the different method.
The space-times in question are of complicated forms and they are not given in the AdS gravity.

In order to simplify the calculations we shall exclusively consider the probe limit, 
when the ratio of the five-dimensional gravitational constant to $U(1)$-gauge field constants will tend to zero.
Due to this limit we take into account $U(1)$-gauge fields living on this fixed background of 
black brane in question. They will satisfy the adequate equations of motion.
The above procedure is widely studied in AdS/CMT approach, e.g., in the case of 
five-dimensional case of $SU(2)$-Yang Mills with magnetic components \cite{amm11}.

The ansatz for static five-dimensional topological black brane with planar symmetry is of the form
\be
ds^2 = - f(r)dt^2 + \frac{dr^2}{f(r)} + r^2 (dx^2 + dy^2 + dz^2).
\ee
The $R_{xx}$ term of Einstein scalar field gravity equations of motion will reveal that
\be
f(r) = \frac{r^2}{2 L^2} - \frac{m}{r^2},
\ee
where $m$ is constant.  The Hawking temperature is provided by the expression
\be
T = \frac{1}{4 \pi}~f'(r) \mid_{r \rightarrow r_h} = \frac{1}{2 \pi} ~r_h.
\ee 
In our consideration   the radius $L$ of the AdS spacetime we set equal to one.

\section{Kinetic and transport coefficients}\label{sec;kin-tr}
{ With three currents $\bf{J_F,~J_B,~\tQ}$ and three vector fields 
$\mathbf{E_F}=(E_F^x,~E_F^y),~\mathbf{E_B}=(E_B^x,~E_B^y),$ and ~$\mathbf{\xi}=(\xi_x,~\xi_y)$, 
where $\mathbf{E_F}$ and $\mathbf{E_B}$ are interpreted as electric fields in sectors $F$ and $B$ 
respectively, while $\mathbf{\xi}=-\nabla T/T$
represents thermal force due to the temperature gradient, one defines the matrix of kinetic
coefficients
\begin{eqnarray}
& \left(
\begin{array}{c}
{J_{(F)}^i}\\
{J_{(B)}^i}\\
{\tQ^i}\\
\end{array}
\right)
=
& \left(\begin{array}{ccc}
 {\sigma{^i_j}_{(FF)}} &  {\sigma{^i_j}_{(FB)}} & {\alpha{^i_j}_{(F)}}T  \\
 {\sigma{^i_j}_{(BF)}} & {\sigma{^i_j}_{(BB)}} &{\alpha{^i_j}_{(B)}}T  \\
 {\alpha{^i_j}_{(F)}}T & {\alpha{^i_j}_{(B)}}T &\ {\kappa_0{^i_j}}T  \\
 \end{array}
 \right)
\left(
\begin{array}{c}
{E^j_{(F)}}\\
 {E^j_{(B)}}\\
 {\xi^j}\\
\end{array}
\right)
\label{ji-vs-fieldsi}
 \end{eqnarray}
with $i,j=x,y$ and obvious definitions $\sigma_{(ab)}^{ij}=\frac{\partial J^i_{(a)}}{\partial E_{j (b)}}$ 
of various conductances $\sigma_{(ab)}^{ij}$ with $a,~b=F,~B$, thermoelectric components
$\alpha_{(a)}^{ij}=\frac{\partial J^i_{(a)}}{\partial \xi_j}$ and $\kappa_{0}^{ij}=\frac{\partial \tQ^i}{\partial \xi_j}$.
Using the expressions (\ref{jiF})-(\ref{qui}) and definitions (\ref{def-a})-(\ref{def-h}) we find the explicit values of the kinetic coefficients
\ben 
\sigma_{(FF)}^{jk} &=&  \frac{\p J^j_{(F)}(\infty)}{{\p E_k}} \\  \nonumber
&=& r_h~ \delta^{jk}\left[1  + \frac{2\tQ_F(2 \tQ_F A +B C r_h)+(B+\alpha B_{add}/2)(2\tQ_F C+BAr_h)r_h}{ r_h^3~ (A^2 + C^2)}\right] \\ \nonumber
&-& \ep^{jk}\left[\frac{2\tQ_F(2\tQ_F C+BAr_h)+(B+\alpha B_{add}/2)(2\tQ_FA+BCr_h)r_h}{r_h^3~ (A^2 + C^2)}\right] ,
\een
\ben 
\sigma_{(FB)}^{jk} &=&  \frac{\p J^j_{(F)}(\infty)}{{\p B_k}} \\  \nonumber
&=& r_h~ \delta^{jk}\left[\frac{\alpha}{2} + \frac{2\tQ_F(2 \tQ_B A +B_{add} C r_h)
+(B+\alpha B_{add}/2)(2\tQ_BC+B_{add}Ar_h)r_h}{ r_h^3~ (A^2 + C^2)}\right] \\ \nonumber
&-& \ep^{jk}\left[\frac{2\tQ_F(2\tQ_B C+B_{add}Ar_h)+(B+\alpha B_{add}/2)(2\tQ_B A+B_{add}Cr_h)r_h}{r_h^3~ (A^2 + C^2)}\right] ,
\een
\ben 
\sigma_{(BF)}^{jk} &=&  \frac{\p J^j_{(B)}(\infty)}{{\p E_k}} \\  \nonumber
&=& r_h~ \delta^{jk}\left[\frac{\alpha}{2} + \frac{2\tQ_B(2 \tQ_F A +B C r_h)
+(B_{add}+\alpha B/2)(2\tQ_F C+BAr_h)r_h}{ r_h^3~ (A^2 + C^2)}\right] \\ \nonumber
&-& \ep^{jk}\left[\frac{2\tQ_B(2\tQ_F C+B Ar_h)+(B_{add}+\alpha B/2)(2\tQ_F A+B Cr_h)r_h}{r_h^3~(A^2 + C^2)}\right] ,
\een
\ben
\sigma_{(BB)}^{jk} &=&  \frac{\p J^j_{(B)}(\infty)}{{\p B_k}} \\  \nonumber
&=& r_h~ \delta^{jk}\left[1  + \frac{2\tQ_B(2 \tQ_B A + B_{add} C r_h)+(B_{add}+\alpha B/2)(2\tQ_BC + B_{add}Ar_h)r_h}{r_h^3~ (A^2 + C^2)}\right] \\ \nonumber
&-& \ep^{jk}\left[\frac{2\tQ_B(2\tQ_B C+B_{add}Ar_h)+(B_{add}+\alpha B/2)(2\tQ_BA+B_{add}Cr_h)r_h}{r_h^3~ (A^2 + C^2)}\right].
\een
In the similar way one arrives at
\be
\alpha_{(F)}^{jk}=\frac{8\pi r_h}{(A^2+C^2)}[\delta^{jk}(2\tQ_FA-(B+\alpha B_{add}/2)C)-\ep^{jk}(2\tQ_FC+(B+\alpha B_{add}/2)A)],
\ee
and
\be
\alpha_{(B)}^{jk}=\frac{8\pi r_h}{(A^2+C^2)}[\delta^{jk}(2\tQ_BA-(B_{add}+\alpha B/2)C)-\ep^{jk}(2\tQ_BC+(B_{add}+\alpha B/2)A)].
\ee
Similarly one finds the kinetic coefficient $\kappa_0$ describing the heat flow under the
temperature bias. It reads
\be
\kappa_0^{jk} = \frac{1}{T}~\bigg( \frac{\p \tQ^j_{(F)}(\infty)}{\p \xi_k} \bigg)=
\frac{64 \pi^2~T~r_h^3~(A~\delta^{jk} - C~\ep^{jk})}{A^2 + C^2}.
\ee 

It has to be noted that the knowledge of the above kinetic coefficients 
is enough to define relevant transport coefficients.

In particular the conductivity tensor $\sigma^{i_j}$ of the material described by the two current model
requires $E_F=E_B=E$, with $E$ being an electric field acting on both electrons 
$Q_{(F)}=-n_e e$ and holes $Q_{(B)}=+n_h e$, 
with $e$ denoting an electric charge. This directly leads to the tensor of total conductivity of the system 
\be
\sigma^{ij}=\sum_{a,b}^{F,B}\sigma_{(ab)}^{ij}.
\ee
The resistivity tensor $\rho^{ij}$ is just the inverse of the conductivity one.
To calculate thermoelectric  tensor one uses standard definition resulting from 
the equation (\ref{ji-vs-fieldsi}) with the 
auxiliary
conditions $J^i_{(F)}=J^i_{(B)}=0$. They allow 
one to find the relations among the fields ${\bf E}_F={\bf E}_B$ and temperature gradient $\nabla T$.
Namely one has that the following relation is satisfied:
\begin{eqnarray}
& \left(
\begin{array}{c}
0\\
0\\
 \end{array}
\right)
=
& \left(\begin{array}{cc}
 {\sigma{^i_j}_{(FF)}} &  {\sigma{^i_j}_{(FB)}}   \\
 {\sigma{^i_j}_{(BF)}} & {\sigma{^i_j}_{(BB)}}   \\
 \end{array}
 \right)
\left(
\begin{array}{c}
{E^j_F}\\
{E^j_B}\\
\end{array}
\right)+
\left(
\begin{array}{c}
{\alpha^i_{j (F)}}\\
{\alpha^i_{j (B)}}\\
\end{array}
\right)\xi^j.
\label{0-vs-fields}
 \end{eqnarray}  
The final expression can be easily found, but 
we do not present its exact form here. 
It defines thermoelectric constant for each of the fields, i.e. $S_{F,B}^{ij}$. 
Our interests are focused on the
definition which requires that the sum of the currents ${\bf J_F}+{\bf J_B}$ vanishes for ${\bf E_F}={\bf E_B}$, 
as one defines the transport coefficients for the semiconductor with two currents. 
The  semiconducting model leads to the definition of the thermoelectric 
tensor $S^{ij}=(\sigma^{-1})^{il}\alpha_l^j$, where $\sigma^{ij}=\sum_{a,b}^{F,B}\sigma_{(a,~b)}^{ij}$ 
and $\alpha^{ij}=\sum_a^{F,B} \alpha_{(a)}^{ij}$.

In our paper, we assumed that the magnetic field is directed along the $z$-axis, so we are
not be able to discuss the effects connected with $\mathbf{E}\cdot\mathbf{B}$ or $\mathbf{\xi}\cdot\mathbf{B}$ 
terms, sometimes called 'axial-gravitational anomalies' \cite{gooth2017} .

{The analysis of the results show that both the magneto-conductance and magneto-resistance are
sensitive functions of the holographic dissipation parameter $\beta$, which on the condensed matter side
we interpret as the inverse mobility $\mu$ of carriers. Precisely, in the studied 3d system one can
identify 
\be
\mu^2=\frac{1}{12~\beta^2~ r_h^2}.
\label{mu-vs-beta}
\ee
This identification allows us to write the magneto-conductance in terms of $\mu B$ product characteristic for the
Drude-Boltzmann approach.   To see this and to answer a natural questions if there are any strong 
coupling features in the obtained formulas and under which
conditions the holographic results reduce to the known classic Boltzmann-like description, we shall rewrite 
some of the kinetic coefficients in terms of ($\mu B$). To make the answer 
clear we shall rewrite the magnetic field dependent Hall component of the conductivity tensor $\sigma^{ij}$ 
in a standard form for 
a single current model and with mobility defined by (\ref{mu-vs-beta}). This leads to the relation
\be
\sigma^{xy}=\frac{8Q_F\mu^2Br_h[(4Q_F^2 +B^2r_h^2)\mu^2+((\mu B)^2+1)]}{[(\mu B)^2+1]^2+16\mu^4Q_F^2B^2r_h^2},
\label{one-curr-hall}
\ee
which is the odd function of the field $B$ and contains corrections to the Drude result
\be
\sigma_D^{xy}=\frac{\sigma_0~\mu~ B}{(\mu B)^2+1},
\label{drude}
\ee
where in the standard notation $\sigma_0=ne\mu$. In order to obtain the Drude-like expression we have
to neglect the second term in the denominator of (\ref{one-curr-hall}) and two terms in the nominator.  
All these terms provide corrections to the  expression (\ref{drude}).}

The holographic analysis shows that the  mobility is inversely proportional to temperature if
one identifies $r_h\approx T$. This is approximate relation valid in the {\it probe limit} only.
To calculate and analyze the charge carrier density dependence of the thermal conductivity of 
graphene with no external magnetic field the authors \cite{seo2017} have diagonalized 
the full matrix (which, in our notation, contained only 
elements $\sigma_{(FF) (BB)}^{xx}$) and got a good agreement with experiment. 
We shall follow this strategy in calculations of charge and thermal conductivity, as well as,
thermoelectric tensors.   We remark in passing that the calculations 
of the thermal conductivity of DSM can be carried out in a full analogy to graphene.
In general thermal conductivity tensor  $\kappa^{ij}$ is defined as 
\be
\tQ^i=-\kappa^{ij}~T~\xi_j,
\ee
under the condition of no current flows in the system.
We shall present and discuss this and other transport parameters 
dependence on the magnetic field in the following section. To calculate 
this parameter we shall require vanishing of both currents as expressed in the equation (\ref{0-vs-fields}).
Calculating all transport coefficients we have assumed $r_h=1$. 

\begin{figure} 
\includegraphics[width=0.45\linewidth]{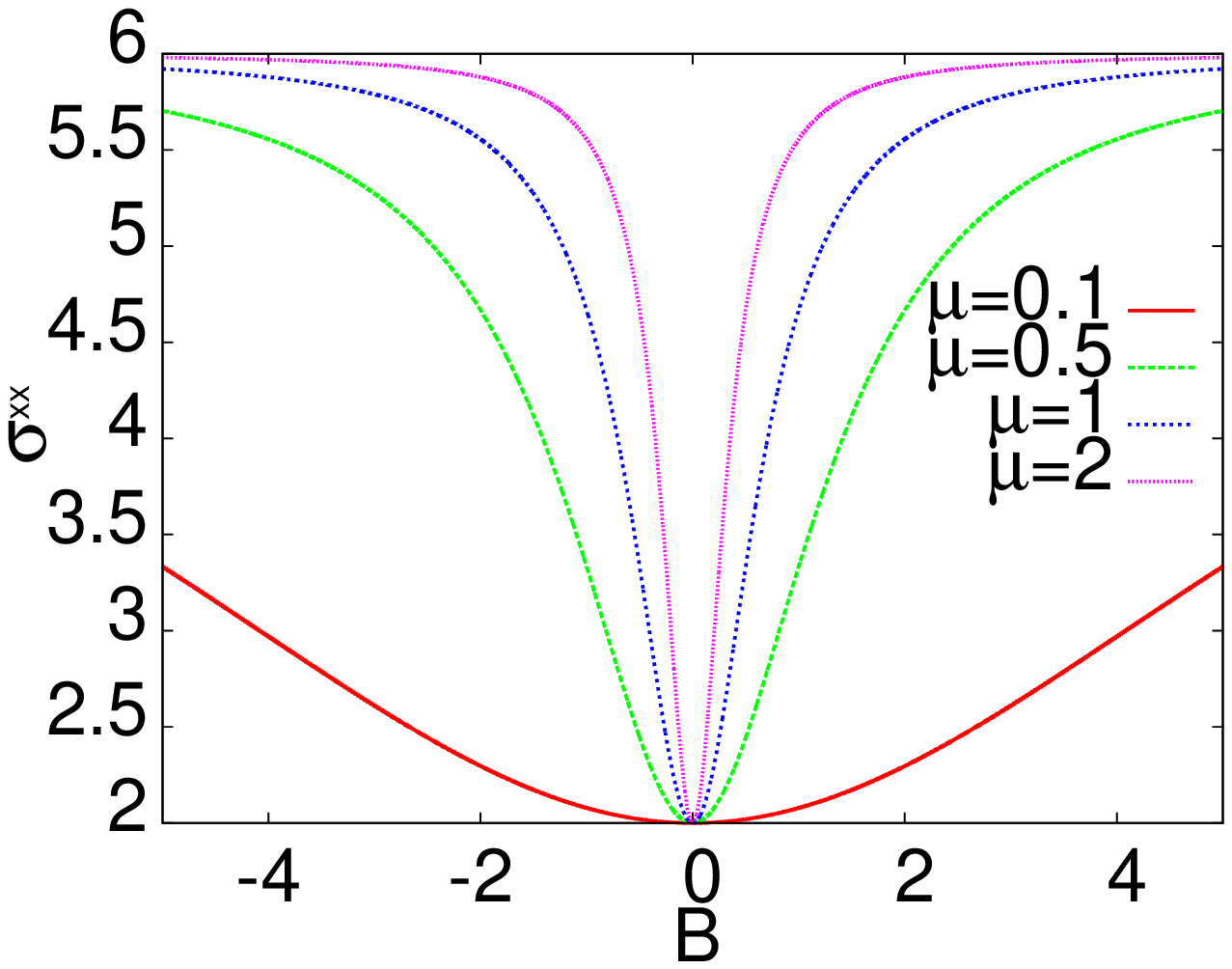} 
\includegraphics[width=0.45\linewidth]{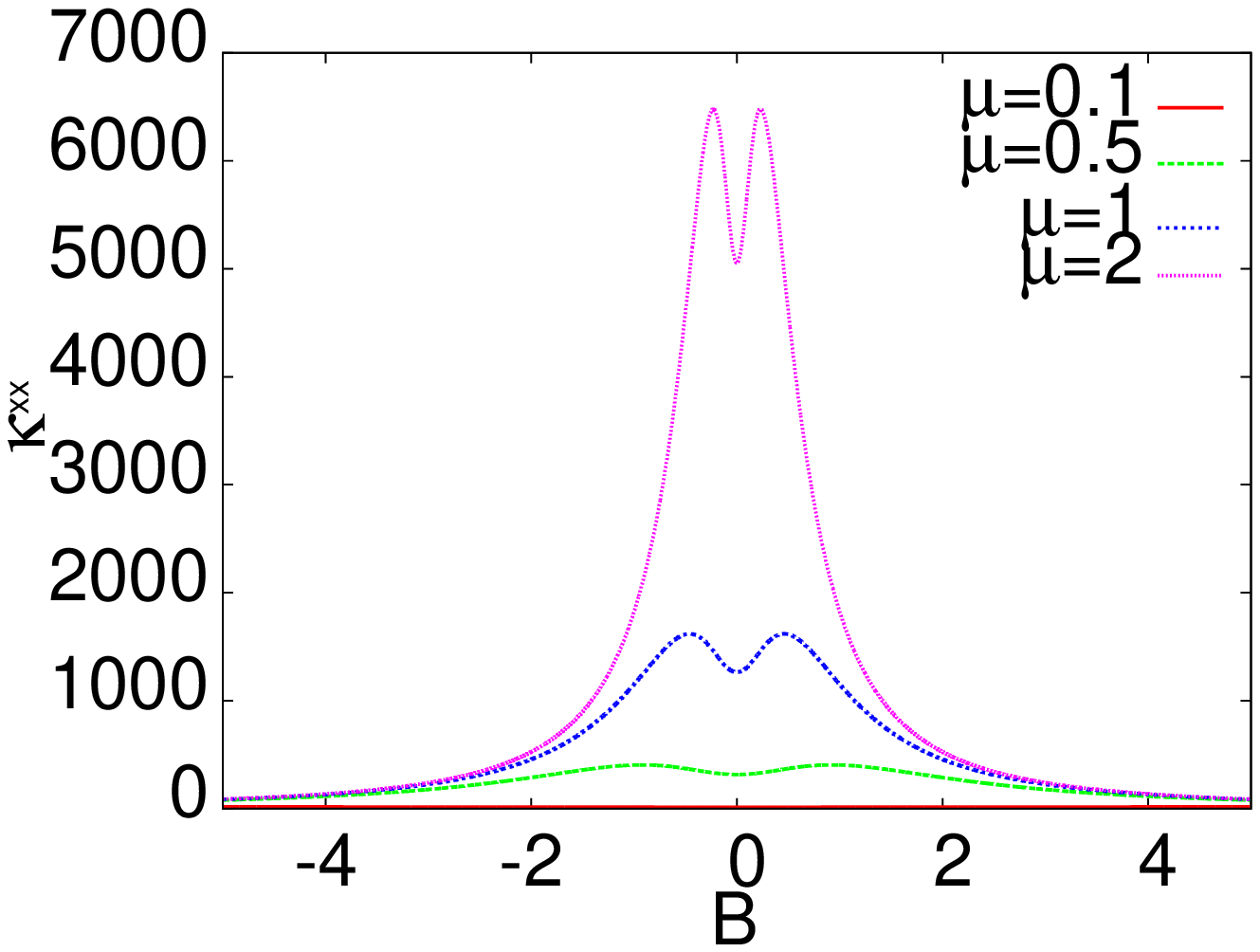} 
\includegraphics[width=0.45\linewidth]{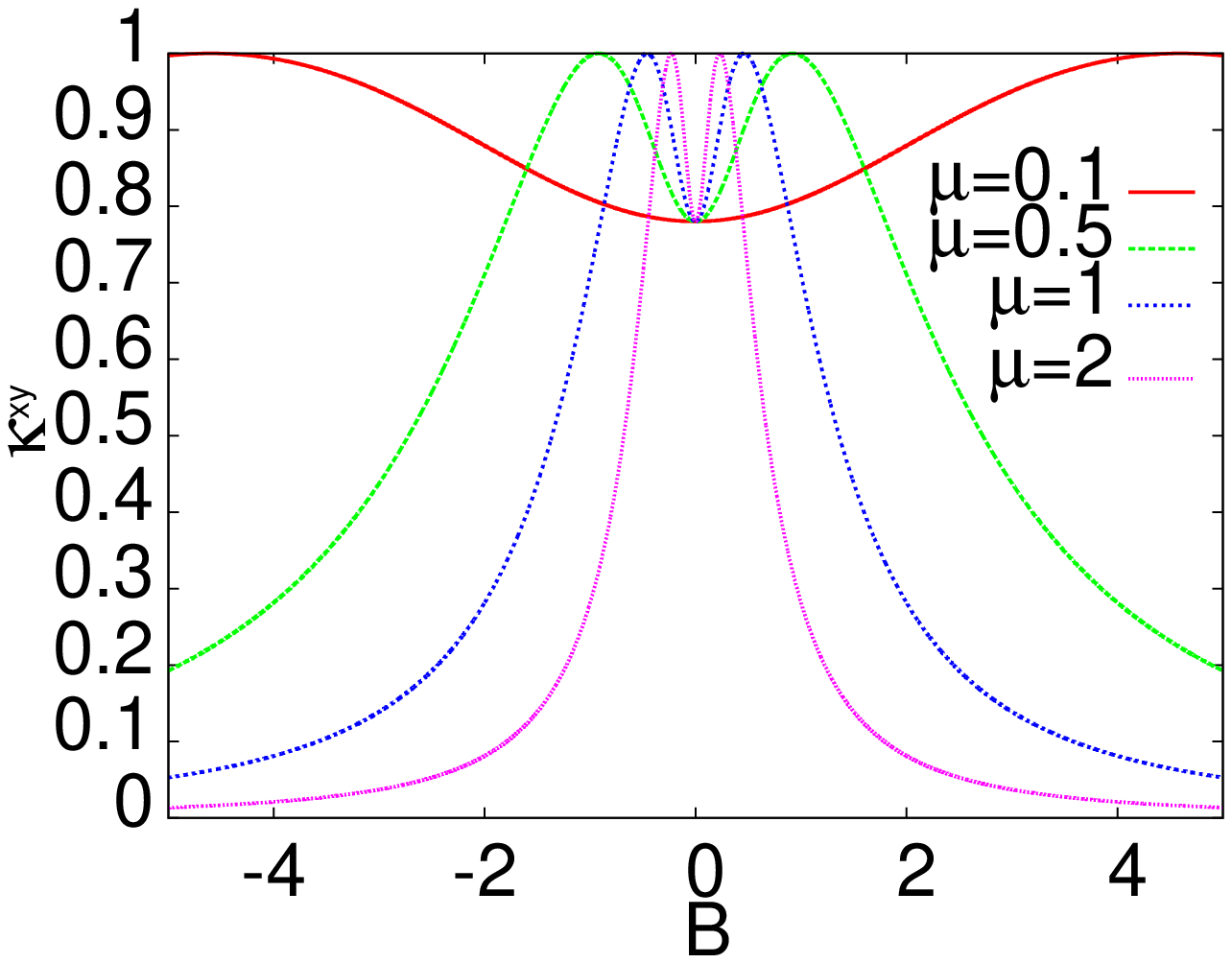} \hspace{1.3cm}
\includegraphics[width=0.45\linewidth]{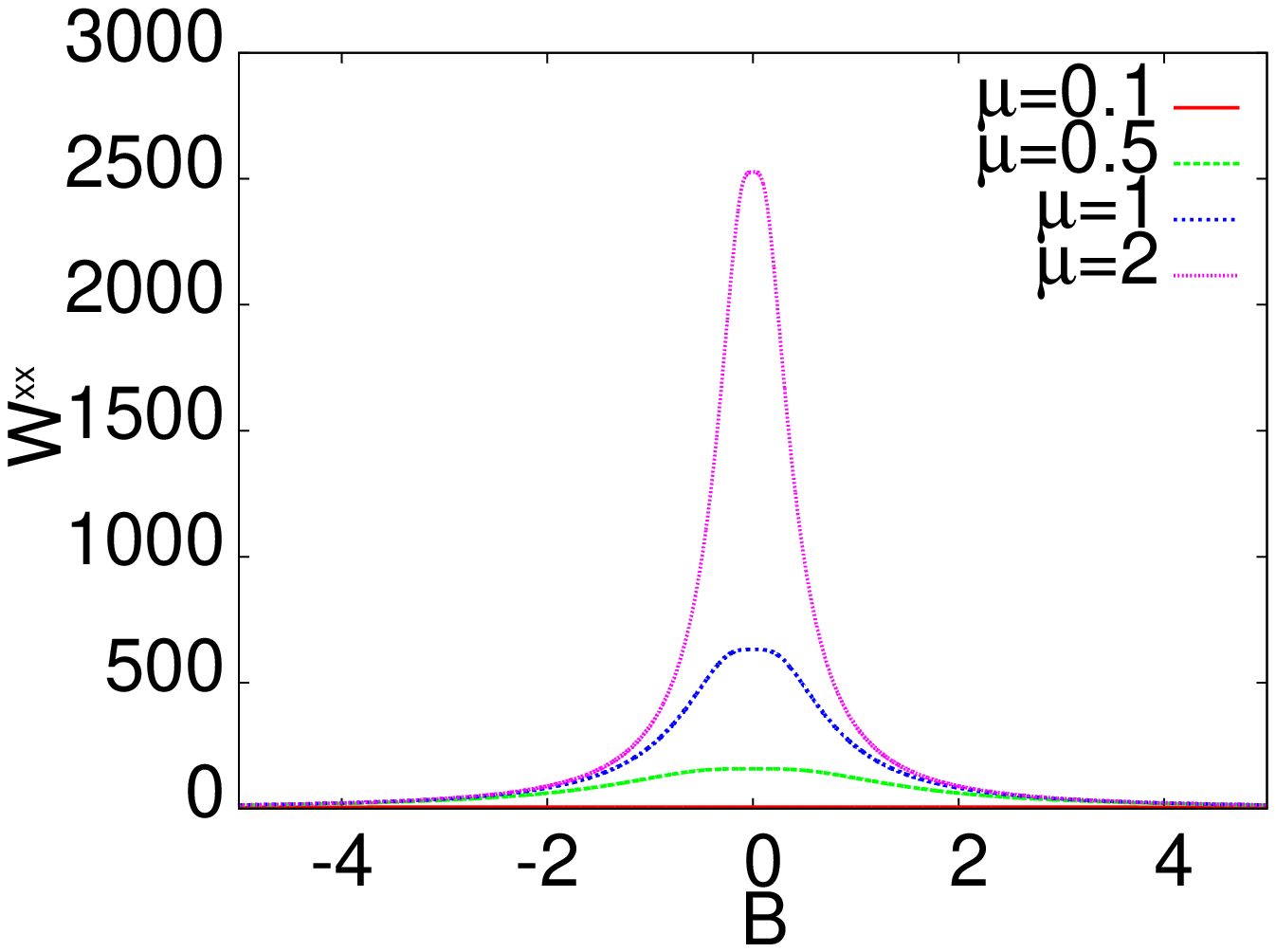}
\caption{ Magnetic field  dependence of the  conductivity  
$\sigma^{xx}$ (upper left panel),  $\kappa^{xx}$ (upper right panel),              
thermal conductivity normalized to its maximal value showing more clearly the effect 
the mobility plays in defining this parameter (lower left panel) and 
Wiedemann-Franz ratio $W^{xx}$ (lower right panel) at the charge neutrality point $n=0$  
for  a few values of the mobility $\mu$.}
\label{fig1-1}
\end{figure}

\section{Experimental verification}
\label{sec:exp}
{Having in mind the kinetic coefficients matrix (\ref{ji-vs-fieldsi}), we calculated magnetic field   dependent transport
coefficients, as was explained in the previous section. Experimentally, the results 
do depend on the sample quality and the studied material. Our model allows to take the sample quality
into account only approximately  by means of the parameter $\beta$ or the mobility $\mu$. The model in question
takes two currents
into account and as mentioned earlier they are connected with electrons and holes.

In general they constitute different particle numbers denoted by $Q_{(F)}$
and $Q_{(B)}$. We shall use the parameter $g$ to characterize their relative contribution. We define $Q=Q_{(F)}+Q_{(B)}$ and introduce
$g$ by requiring that $Q_{(F)} - Q_{(B)}= gQ$ and identifying $Q$ as the effective carrier concentration $n$.
 With this choice, motivated by the previous work on graphene \cite{seo2017} vanishing of $Q$ is accompanied 
by vanishing of both $Q_{(F)}$ and $Q_{(B)}$.  As already noted it 
is rather difficult to change $n$ in $(3+1)$-dimensional materials with Dirac spectrum. At the same time
the actual concentration of carriers is also not equal to zero. That is why we shall concentrate on the
magnetic field dependence of the transport characteristics for the assumed values of $n$. 
To this end we assume $B_{add}=B$ in what follows.
In the equilibrium state and exactly
at the Dirac point, one has that  $Q=0$. The third free parameter of the model is the coupling $\alpha$ 
between the two $U(1)$-gauge fields.
We shall study the effect of both $g$ and $\alpha$ on the magnetic field dependence for various transport coefficients. 

For the perfectly compensated system with $n=0$, all off-diagonal transport coefficients vanish.
This can be seen 
from their definitions, when the effective charge densities go to zero.
On the other hand, the following conclusions can be drawn
from figure \ref{fig1-1}. 
The magneto-resistance (MR) defined as the ratio $MR=(\rho^{xx}(B)-\rho^{xx}(0))/\rho^{xx}(0)$
is always positive if there is no mixing ($g=0$) and no interactions ($\alpha=0$) between currents. 
The minimal value of the conductivity
appears at $B=0$ and does not depend on $\mu$, the effective mobility of carriers. However, the thermal conductivity
strongly increases with the growth of the mobility. Moreover, it has a local minimum for $B=0$ and two maxima at finite
values of the magnetic field. The lower left panel of the figure shows the decrease of the width of the $\kappa^{xx}$ with
increasing of the mobility. The lower right panel of figure \ref{fig1-1} shows the Wiedemann - Franz ratio of this 
compensated system $vs.$ magnetic field. 
It happens that with the growth of $\mu$, the ratio strongly increases.
At the same time, the
width of the curve $W^{xx}(B)$ decreases.

Figures \ref{fig1-2}-\ref{fig1-6} show the diagonal and off-diagonal components of the thermal conductivity,
resistivity,  thermoelectric power and Wiedemann-Franz ratio, respectively,  as a function of 
magnetic field calculated for a small 
but  finite value of the carrier concentration $n=0.1$ and for 
a few values of mobility parameter (as indicated in the figures). 

\begin{figure} 
\includegraphics[width=0.45\linewidth]{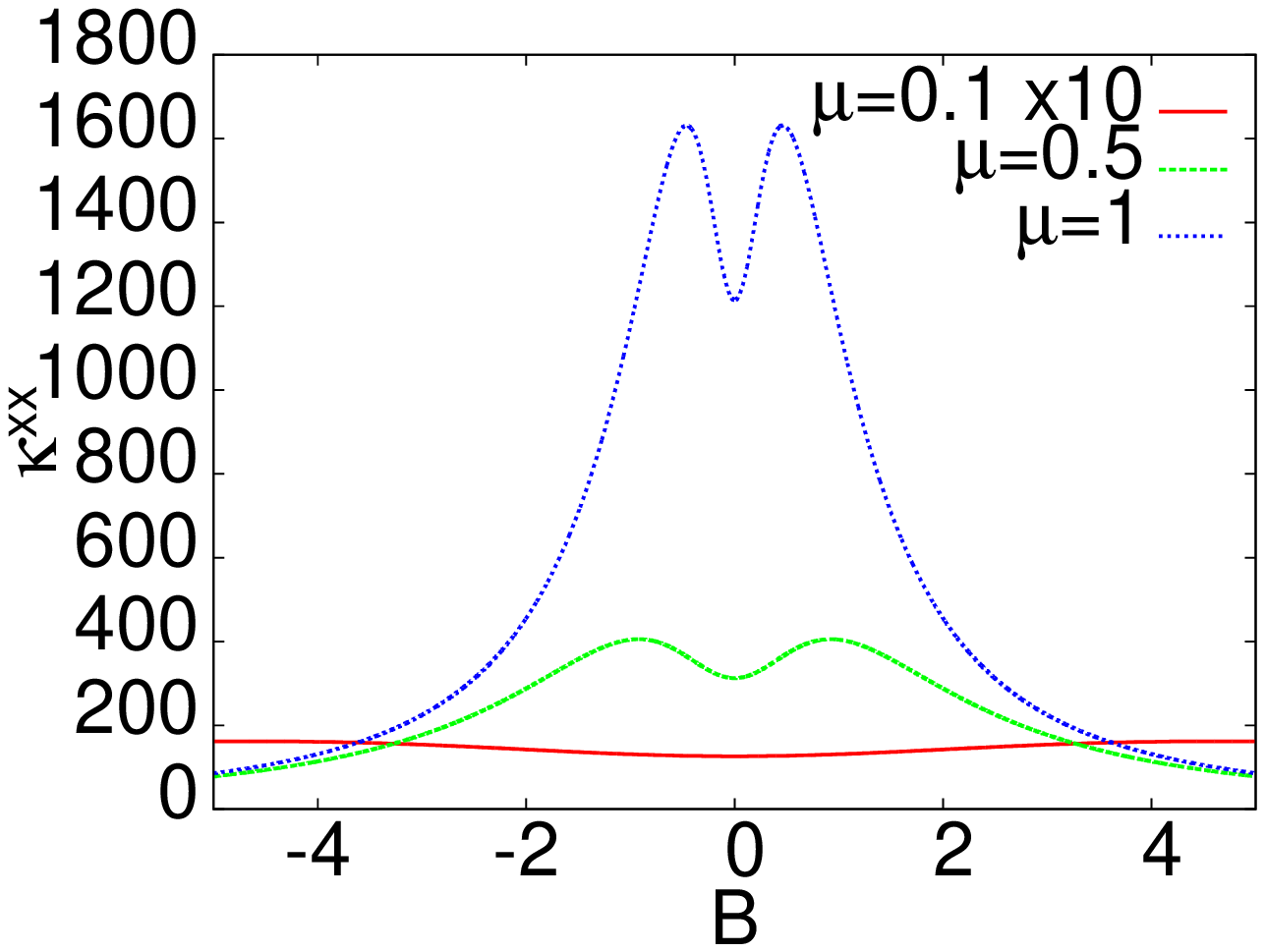} \hspace{1.3cm}
\includegraphics[width=0.45\linewidth]{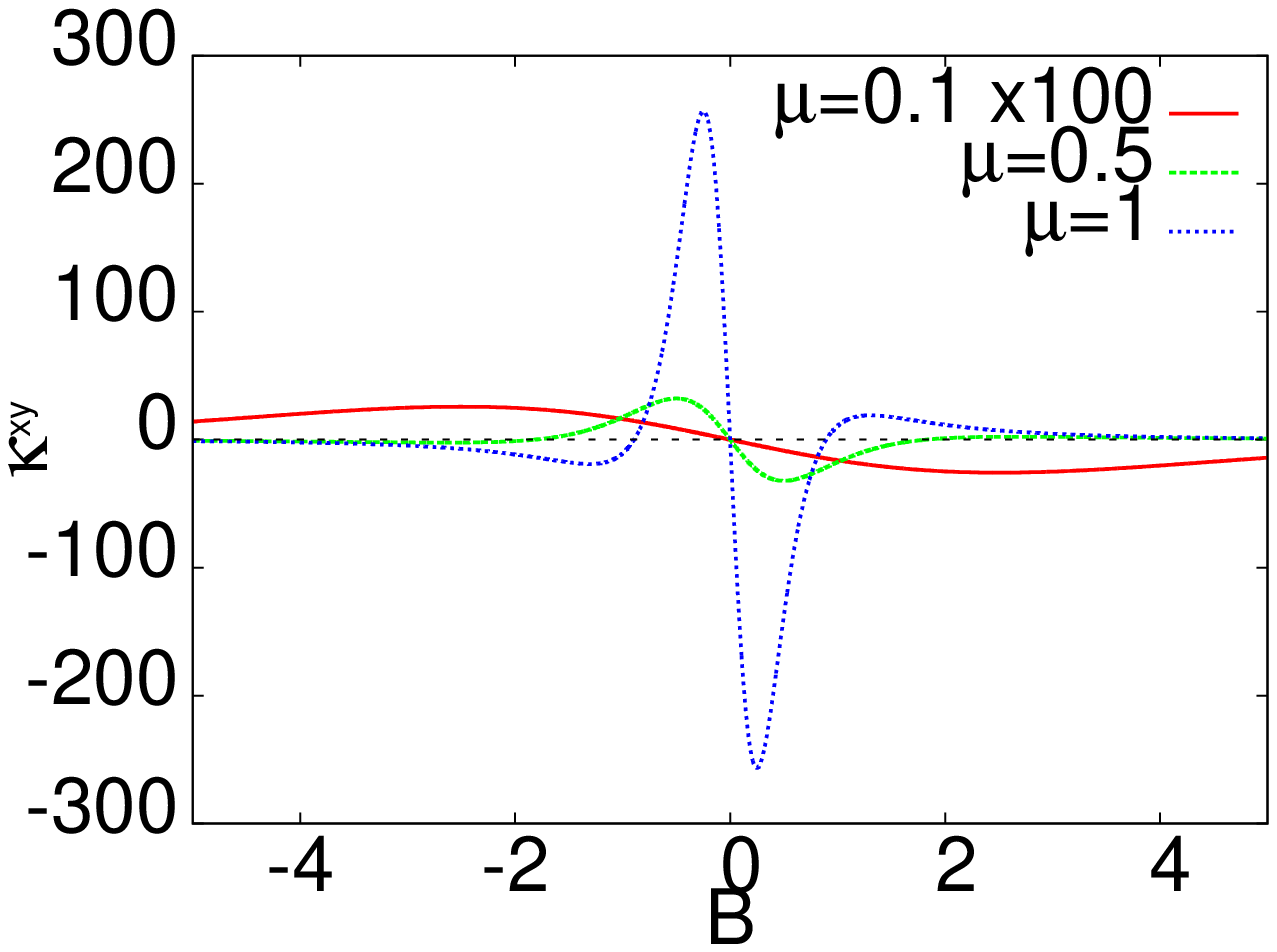}
\caption{ Magnetic field  dependence of the thermal  conductivity  
$\kappa^{xx}$ (left panel) and $\kappa^{xy}$ obtained for  $g=0$, $\alpha=0$ and charge density $n=0.1$.  
Different curves correspond to mobility parameter $\mu=0.1,~ 0.5$ and 1. For better visibility we multiplied
the curve corresponding to lowest mobilities by the numerical factors 10 or 100. }
\label{fig1-2}
\end{figure}
The maximal values of the thermal conductivity tensor, shown in the figure \ref{fig1-2}, 
strongly depend on the mobility $\mu$ also outside the Dirac point. In the figure they are plotted
for $n=0.1$, while the other parameters are fixed to be $g=0$ and $\alpha=0$. 
The component $\kappa^{xx}(B)$ is a symmetric
function of the magnetic field, while $\kappa^{xy}$ is antisymmetric 
with respect to $B$. It turns out that they feature the similar
symmetry properties with respect to $n$, $i.e.$ $\kappa^{xx}(-B,~-n)=\kappa^{xx}(B,~n)$, 
while $\kappa^{xy}(-B,~n)=-\kappa^{xy}(B,~n)$ and
 $\kappa^{xy}(B,~-n)=-\kappa^{xy}(B,~n)$. The diagonal component has a two peak structure
with minimum at $B=0$. The dependence of $\kappa^{xy}$ on the magnetic field 
is rather complicated. It changes sign for $B=0$ and also for finite $B$. The latter point 
depends on the mobility and moves towards lower values with the increasing of $\mu$.

\begin{figure} 
\includegraphics[width=0.45\linewidth]{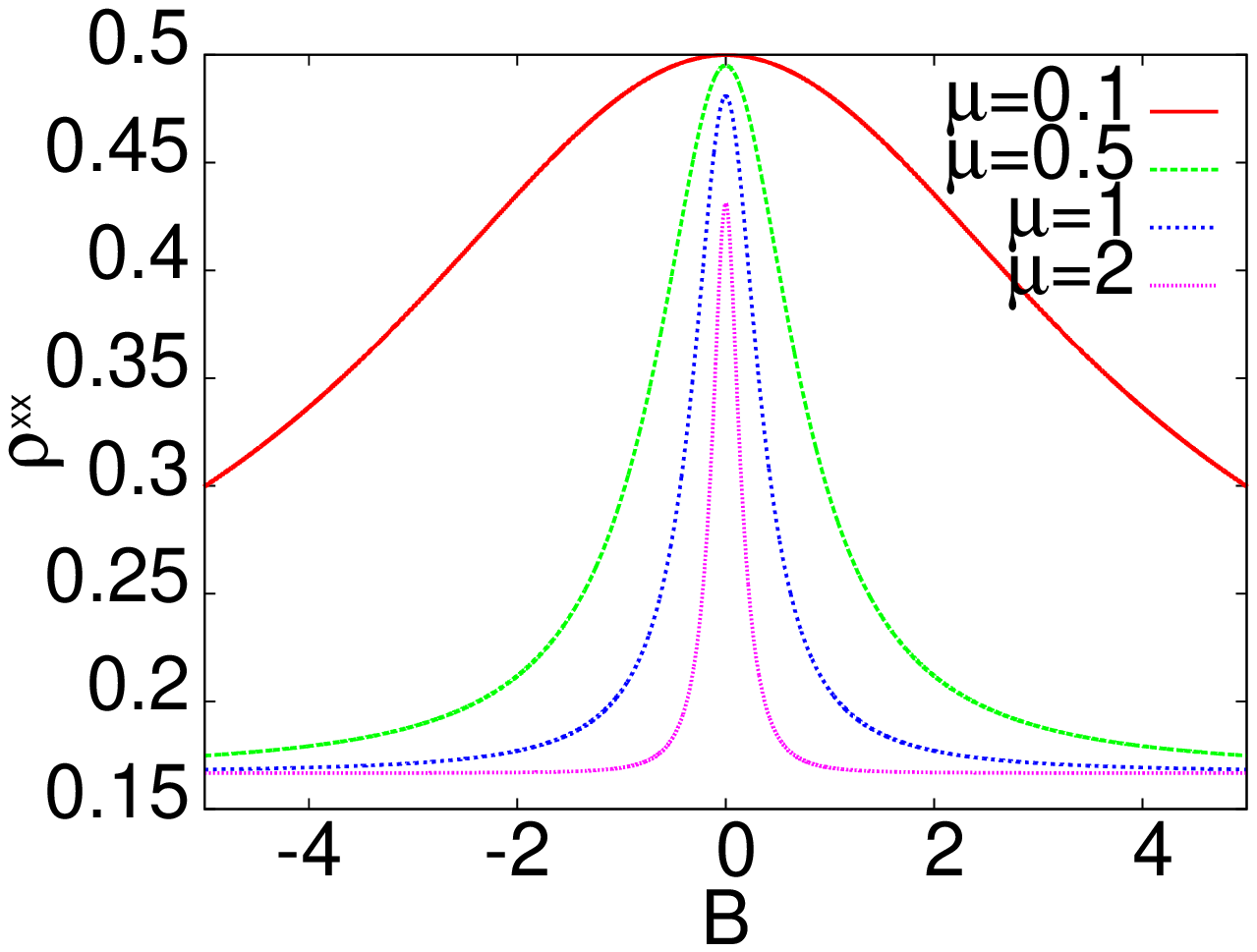} \hspace{1.3cm}
\includegraphics[width=0.45\linewidth]{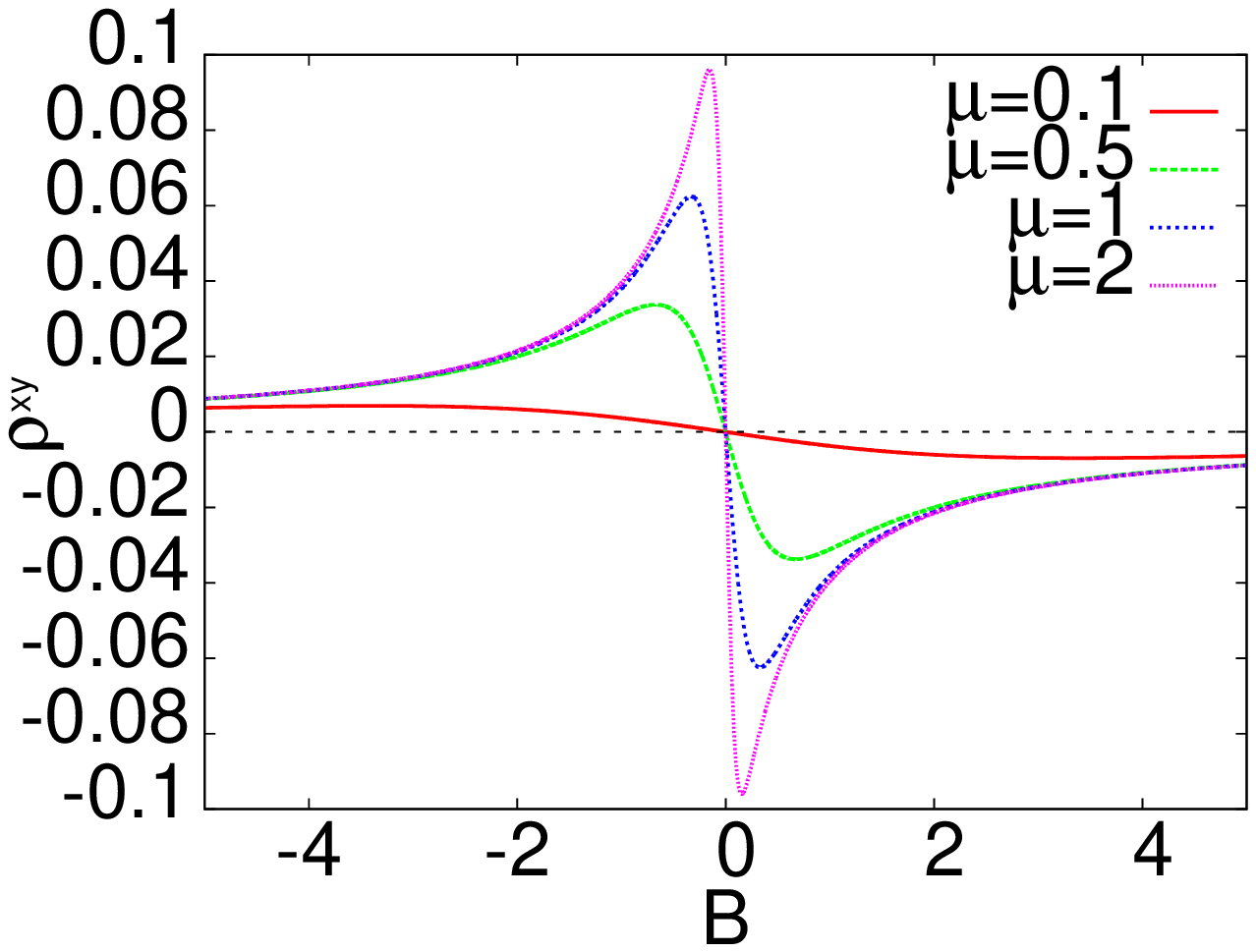}
\caption{ Magnetic field  dependence of the diagonal $\rho^{xx}$ (left panel)  
and off-diagonal $\rho^{xy}$ (right panel) components of the resistivity tensor
for a few values of mobility $\mu$. The other parameters are set to $g=0$, $\alpha=0$  
 and $n=0.1$. Note that the magneto-resistance is negative.}
\label{fig1-3}
\end{figure}

The components of the resistivity tensor are plotted as a function of magnetic field
in the figure \ref{fig1-3}. They possess expected symmetry properties with respect to the
magnetic field. The diagonal resistivity is even function of both $B$ and $n$, while $\rho^{xy}$ is
odd function of $n$ and also of $B$.  
With the increase of $\mu$, the width of the curves $\kappa^{xx}(B)$, 
at half-maximum, narrows. Similarly, the local maxima in $\rho^{xy}$ move towards $B=0$, with the growth of $\mu$.
Such a behavior is observed for all the studied transport parameters and seems to be the general feature
of the holographic approach to strongly interacting particles. In the studied systems really strong 
interactions are expected at and the very close distances to the particle-hole symmetry point. The hydrodynamic
like behavior is related to the phase space restrictions on the possible single particle 
scattering events in systems with linear spectrum.

\begin{figure} 
\includegraphics[width=0.45\linewidth]{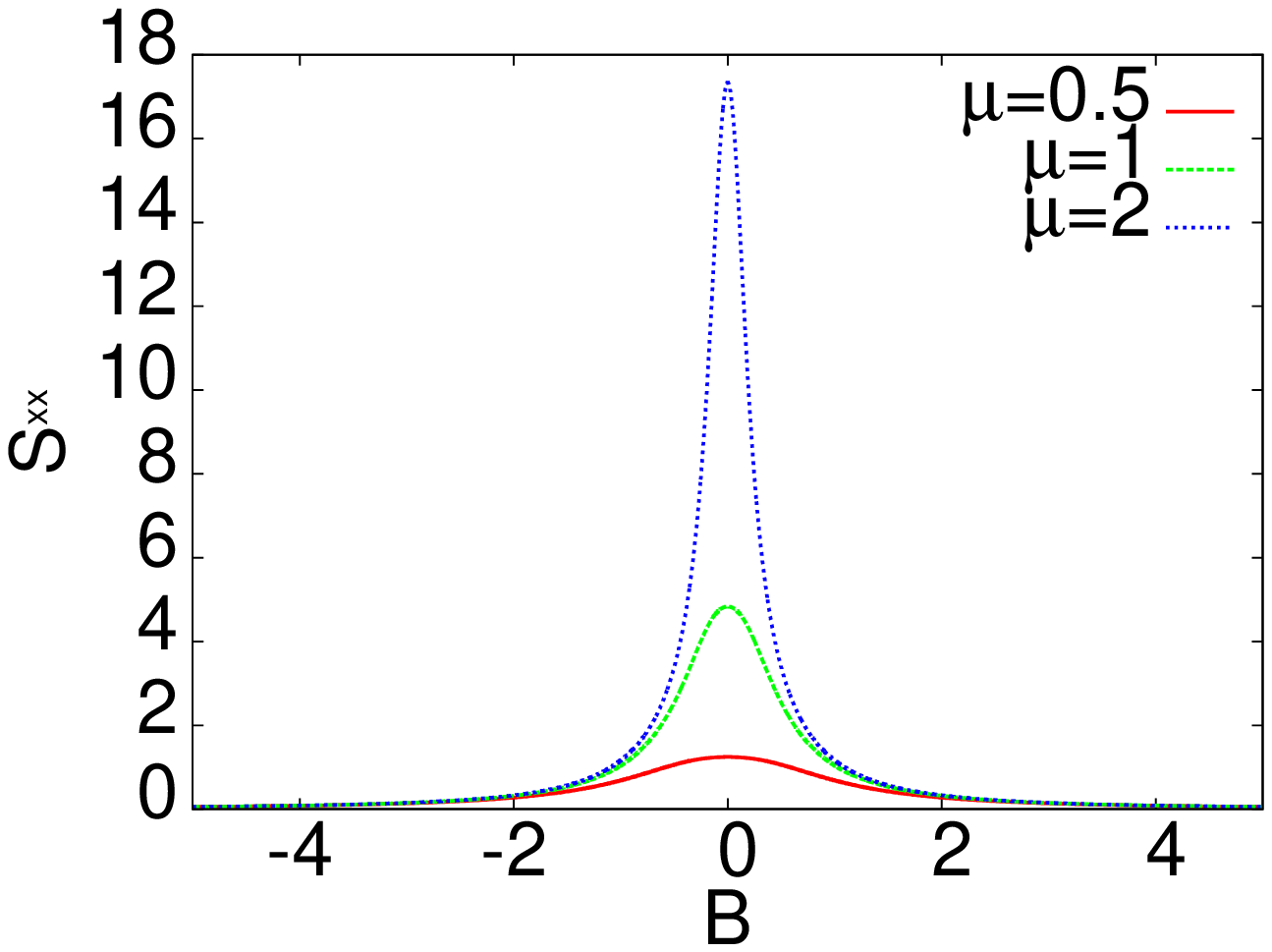} \hspace{1.3cm}
\includegraphics[width=0.45\linewidth]{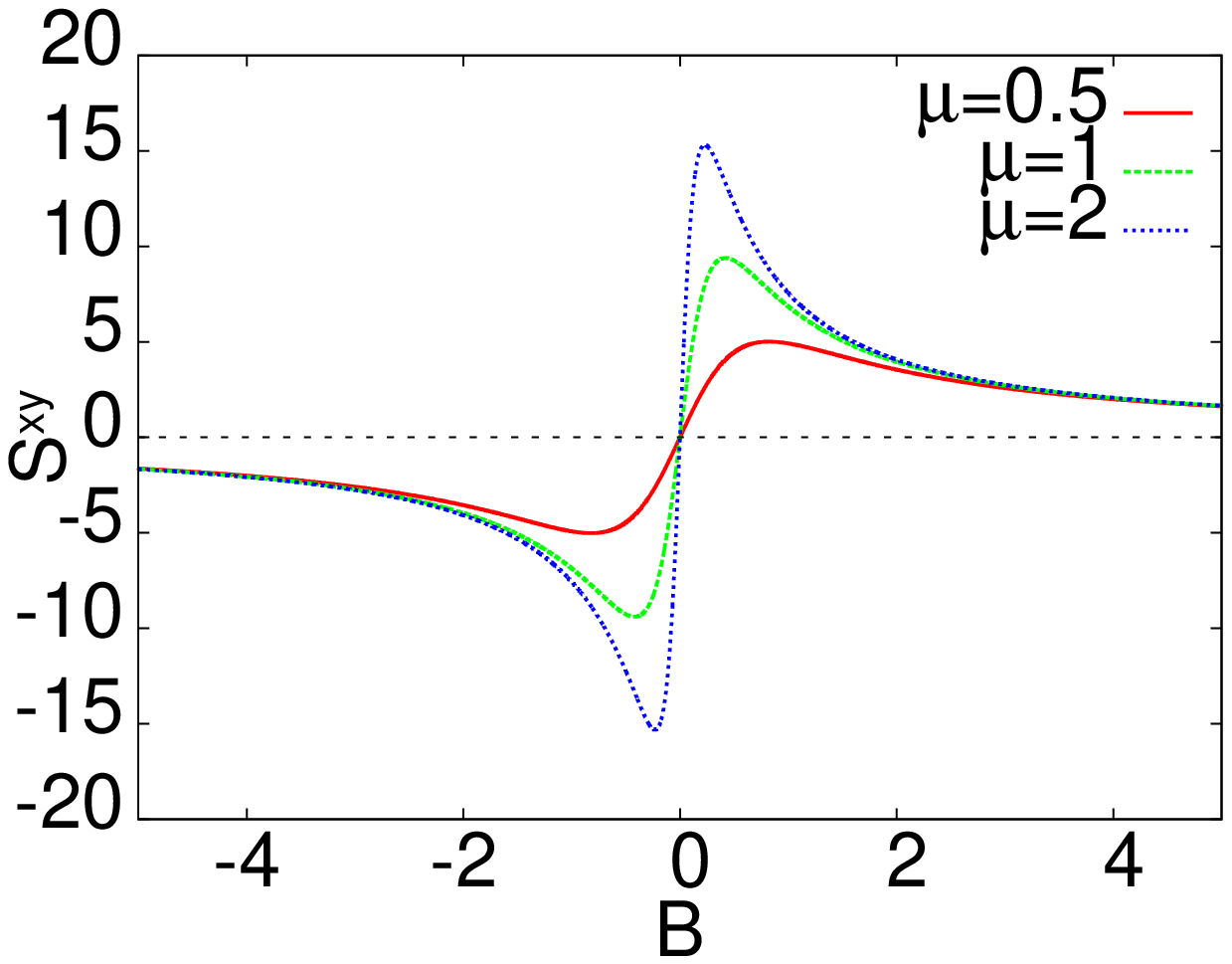}
\caption{ Magnetic field  dependence of the diagonal $S^{xx}$ (left panel)  
and off-diagonal $S^{xy}$ (right panel) components of the t tensor
for a few values of mobility $\mu$. The other parameters are set to $g=0$, $\alpha=0$
and $n=0.1$}
\label{fig1-4}
\end{figure}

The Seebeck and Nernst parameters are given by the respective components of the thermoelectric
tensor. Phenomenologically they are defined as constants of proportionality between the voltage 
appearing in the system in response to the applied temperature gradient, with the auxiliary condition that
the current vanishes. It is known from condensed matter physics that their measurements give
additional information about the spectrum of carriers. In particular, the  Seebeck coefficient can be shown
to depend on the slope of the density of states at the Fermi energy, while the conductivity depends
on the value of the density of states. $S^{xx}$ can also be interpreted as an entropy carried in the
system. Figure \ref{fig1-4} shows the magnetic field dependence of the Seebeck (left panel) and Nernst components
of the thermoelectric tensor. It turns out that $S^{xx}$ is the symmetric function of the magnetic field but
antisymmetric function of the charge density $n$. This is in accord with the standard notion that the Seebeck coefficient 
for electrons is negative and for  holes is positive, and its sign is used to define majority carriers. On the other hand,
the Nernst coefficient is the even function of $n$ and the odd function of magnetic field $B$.

It is customary to define the Wiedemann-Franz ratio $W^{xx}=\kappa^{xx}/(\sigma^{xx}T)$ and
compare its value to the so-called Lorentz constant $L_0$, obtained for the nearly free electron model.
The departures of $W^{xx}$ from $L_0$ ($W^{xx}>L_0$) are considered as signs of strongly interacting particles.
Here we propose slight extension of the Wiedemann-Franz ratio to both diagonal and
non-diagonal components with $W^{xy}=\kappa^{xy}/(\sigma^{xy}T)$. The resulting quantities are plotted 
in the figure \ref{fig1-6}, as the function of magnetic field.
The three curves correspond to the three values of the
mobility and their behavior is in accord with the other transport parameters. Namely,
the magnitude increases with the growth of the mobility $\mu$
and the curves narrow down. The diagonal ratio $W^{xx}(B)$ is positive, while $W^{xy}(B)$ changes 
its sign according to the signs of
$\kappa^{xy}(B)$ and $\sigma^{xy}(B)$. The measurements of these ratios for materials with Dirac spectrum would provide
the additional test of the holographic approach.

\begin{figure} 
\includegraphics[width=0.45\linewidth]{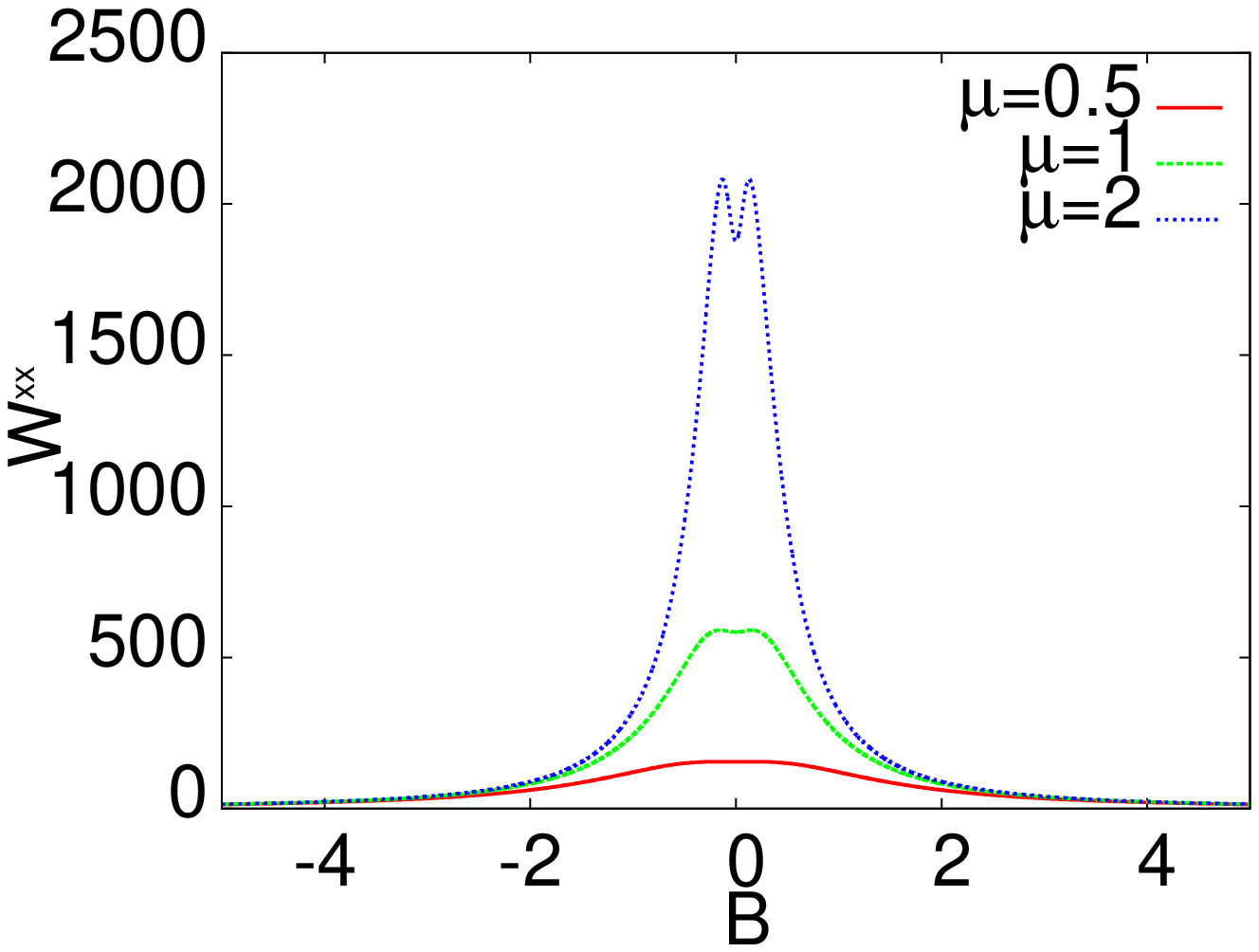} \hspace{1.3cm}
\includegraphics[width=0.45\linewidth]{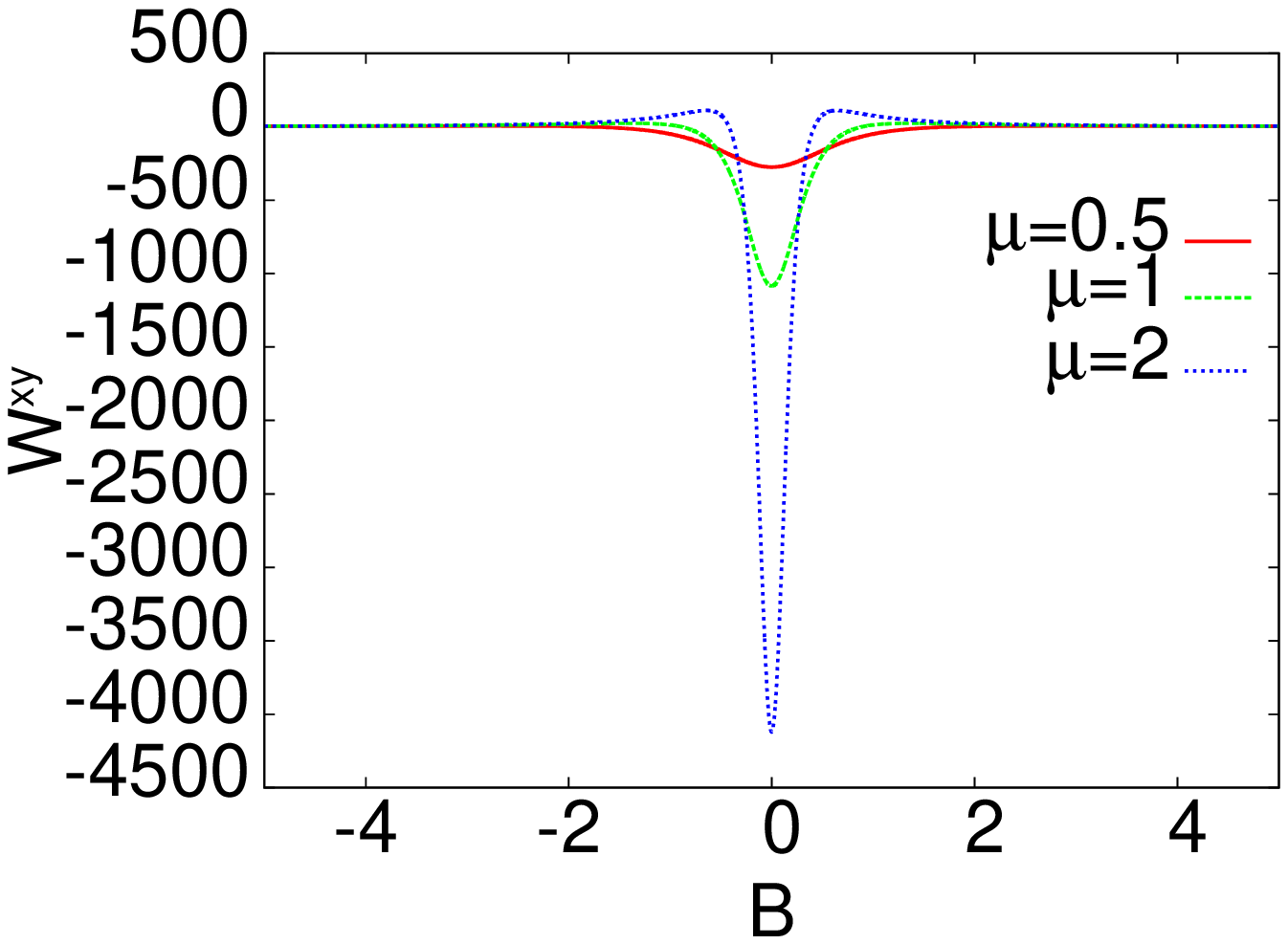}
\caption{ Magnetic field  dependence of the Wiedemann-Franz ratio  
$W^{xx}=\kappa^{xx}/(\sigma^{xx}T)$ (left panel) and $W^{xy}=\kappa^{xy}/(\sigma^{xy}T)$ (right panel) 
for three values of $\mu=0.5,1,2$ and other parameters set to  $n=0.1$, $g=0$  and $\alpha=0$.}
\label{fig1-6}
\end{figure}

\begin{figure} 
\includegraphics[width=0.45\linewidth]{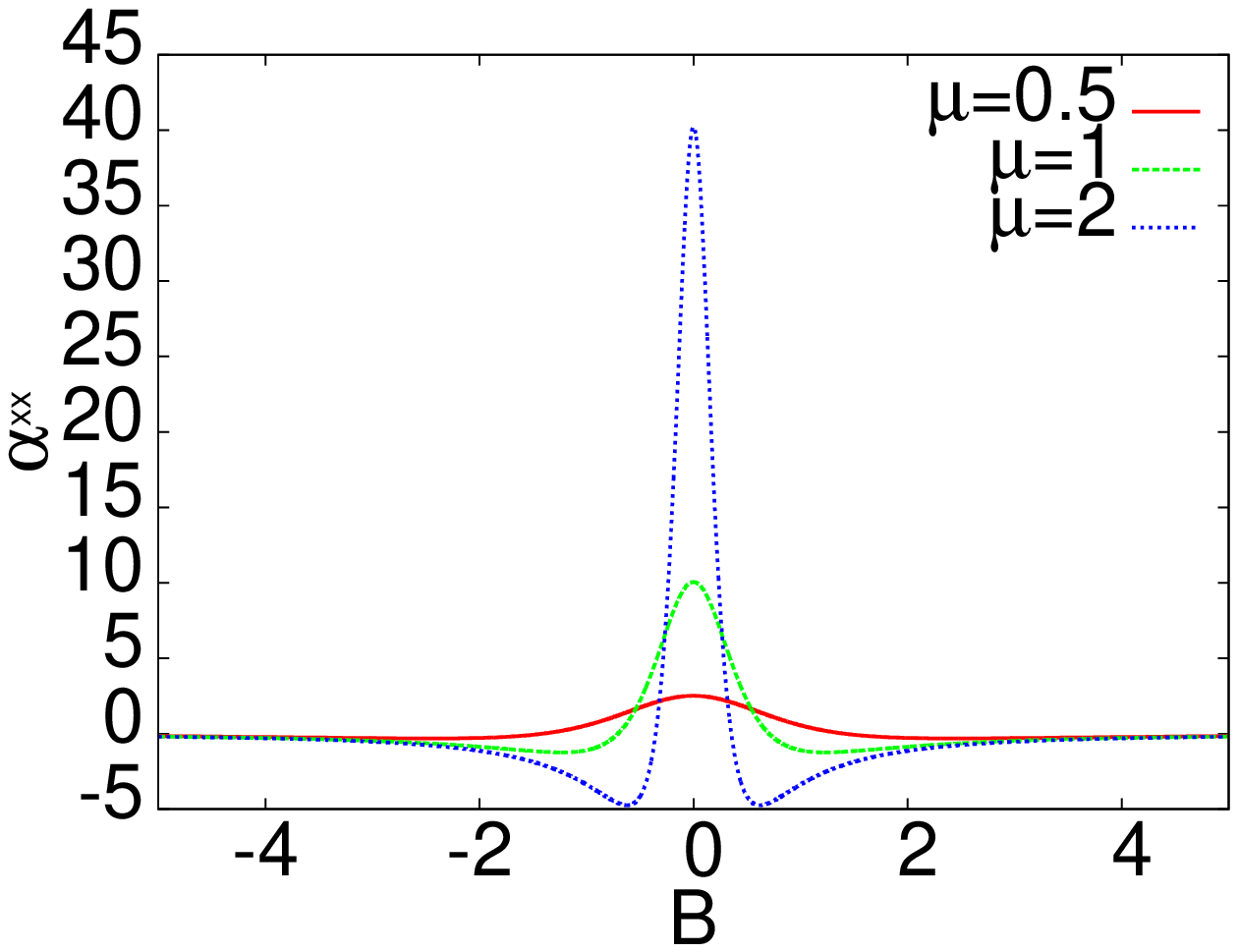} \hspace{1.3cm}
\includegraphics[width=0.45\linewidth]{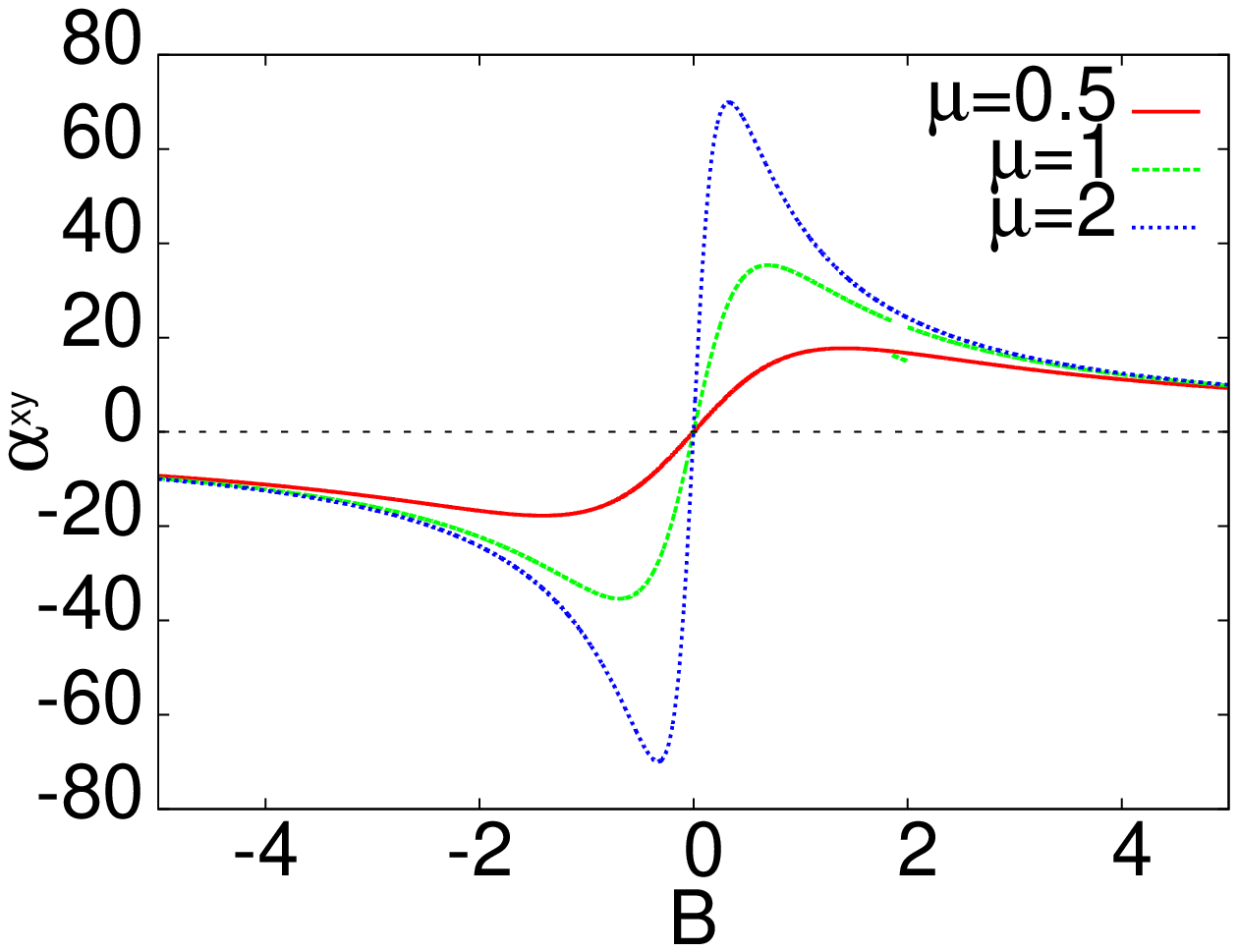}
\caption{ Magnetic field  dependence of the thermal kinetic coefficient 
$ \alpha^{xx} $ (left panel) and $ \alpha^{xy} $ (right panel) 
for three values of $\mu=0.5,1,2$ and other parameters set to  $n=0.1$, $g=0$  and $\alpha=0$.}
\label{fig1-6b}
\end{figure}

Figure \ref{fig1-6b} illustrates the dependence of the bar values of the thermal conductivities
$\alpha^{xx}(B)$ and  $\alpha^{xy}(B)$ on the magnetic field, for a few values 
of the mobility parameter $\mu$. Similarly to the previously discussed transport characteristics,
the increase of $\mu$ leads to the growth of the absolute values of different features 
in these kinetic coefficients.

As mentioned previously, the most often studied 3d system with Dirac spectrum 
is Cd$_3$As$_2$ \cite{liang2015,liang2017,cheng2016}. There have been some experimental 
measurements of magnetic field dependence of the transport parameters of this material. 
It is interesting to note that
the holographically calculated elements of the conductivity  and thermopower tensor 
show resemblance with the experimental data measured for Cd$_3$As$_2$ system \cite{liang2015,liang2017}.
Furthermore, the dependence of the conductivity tensor on the magnetic field features quantitative similarity
to the measurement. This is true for the studied material but also for other
systems \cite{xiong2015}, as it has been reported recently.

We have allowed for the mixing of two currents flowing in the system (parameter $g$) 
and for their interaction (parameter $\alpha$). The effect of $g$ on some of the studied transport
characteristics is shown in the figure \ref{fig1-7}. The figure envisages the effect of $g$
on the magnetic field dependence of both components of the Wiedemann-Franz ratio. The influence
of $g$ is not very big but can be an important factor in the detailed description of experiments. 
The interaction between the fields $\alpha$ also plays a similar role. It changes the maximal values 
of various transport parameters as is illustrated in the figure \ref{fig1-8}, where the longitudinal 
and Hall resistivities are depicted as the function of the magnetic field. This figure is obtained for $g=0$
but quantitatively similar behavior is observed for $g\ne 0$.

\begin{figure} 
\includegraphics[width=0.45\linewidth]{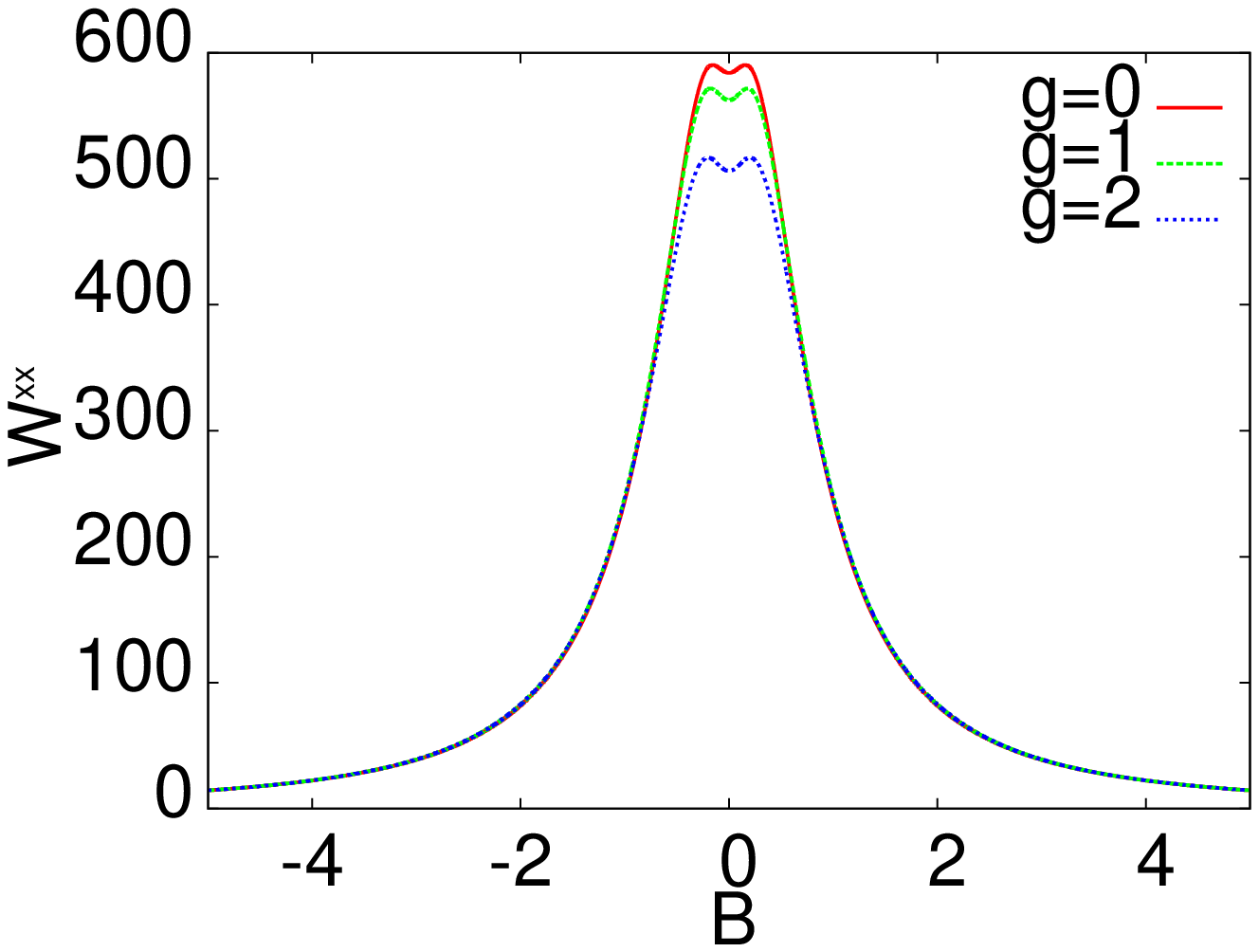} \hspace{1.3cm}
\includegraphics[width=0.45\linewidth]{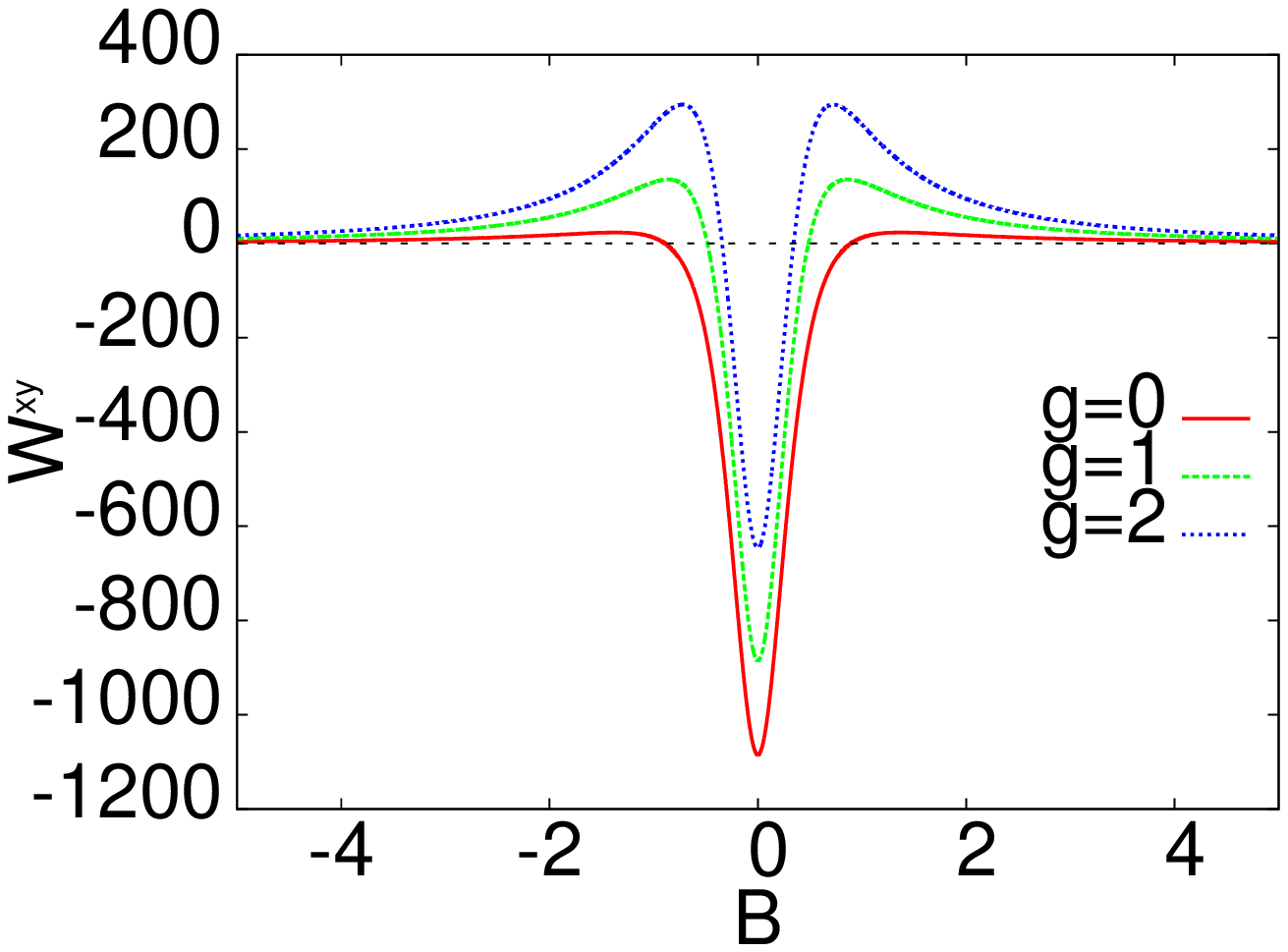}
\caption{ Magnetic field  dependence of the Wiedemann-Franz ratio
$ W^{xx} $ (left panel) and $W^{xy} $ (right panel) 
for three values of $g=0,1,2$ and other parameters set to  $n=0.1$, $g=0$  and $\alpha=0$.}
\label{fig1-7}
\end{figure}
\begin{figure} 
\includegraphics[width=0.45\linewidth]{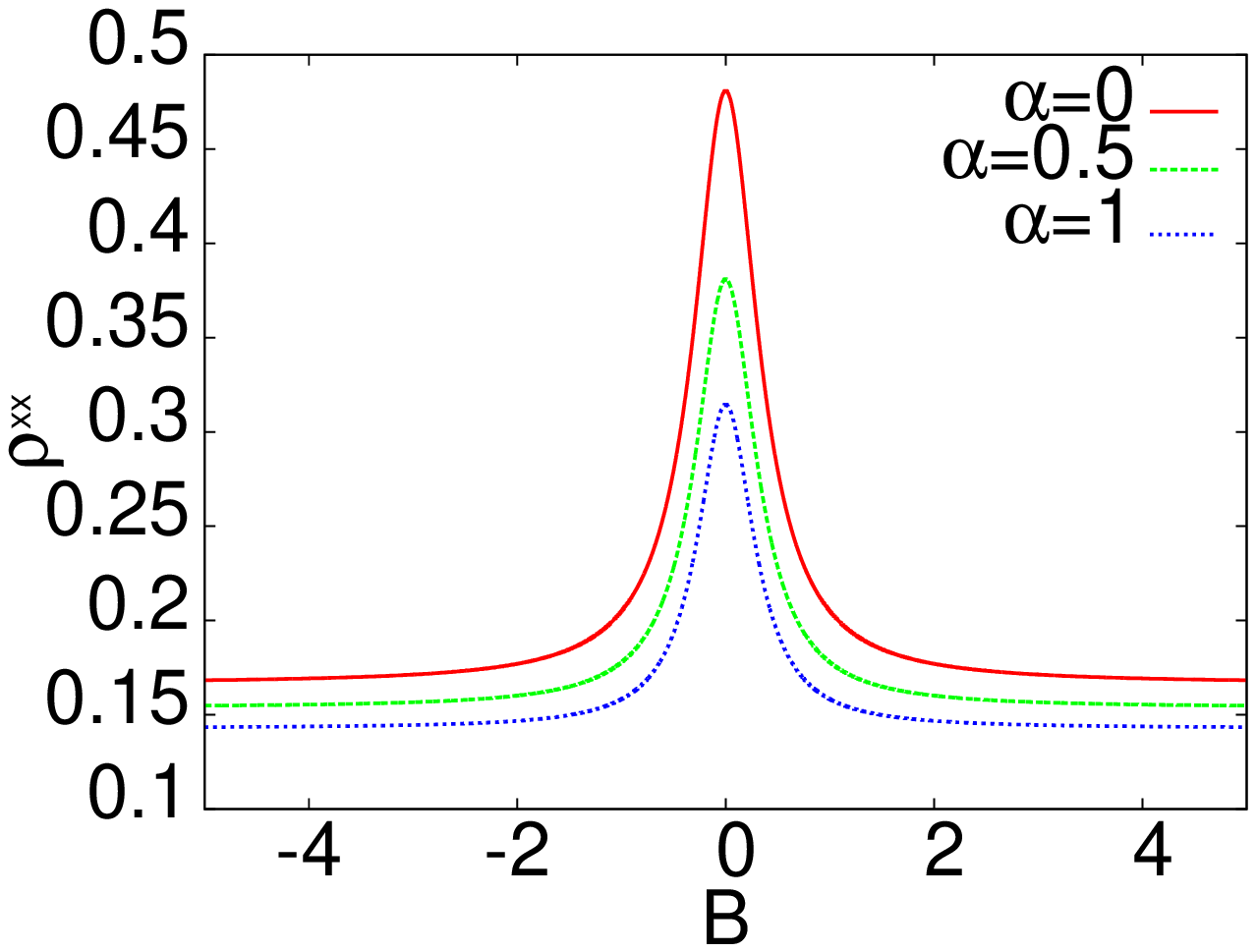} \hspace{1.3cm}
\includegraphics[width=0.45\linewidth]{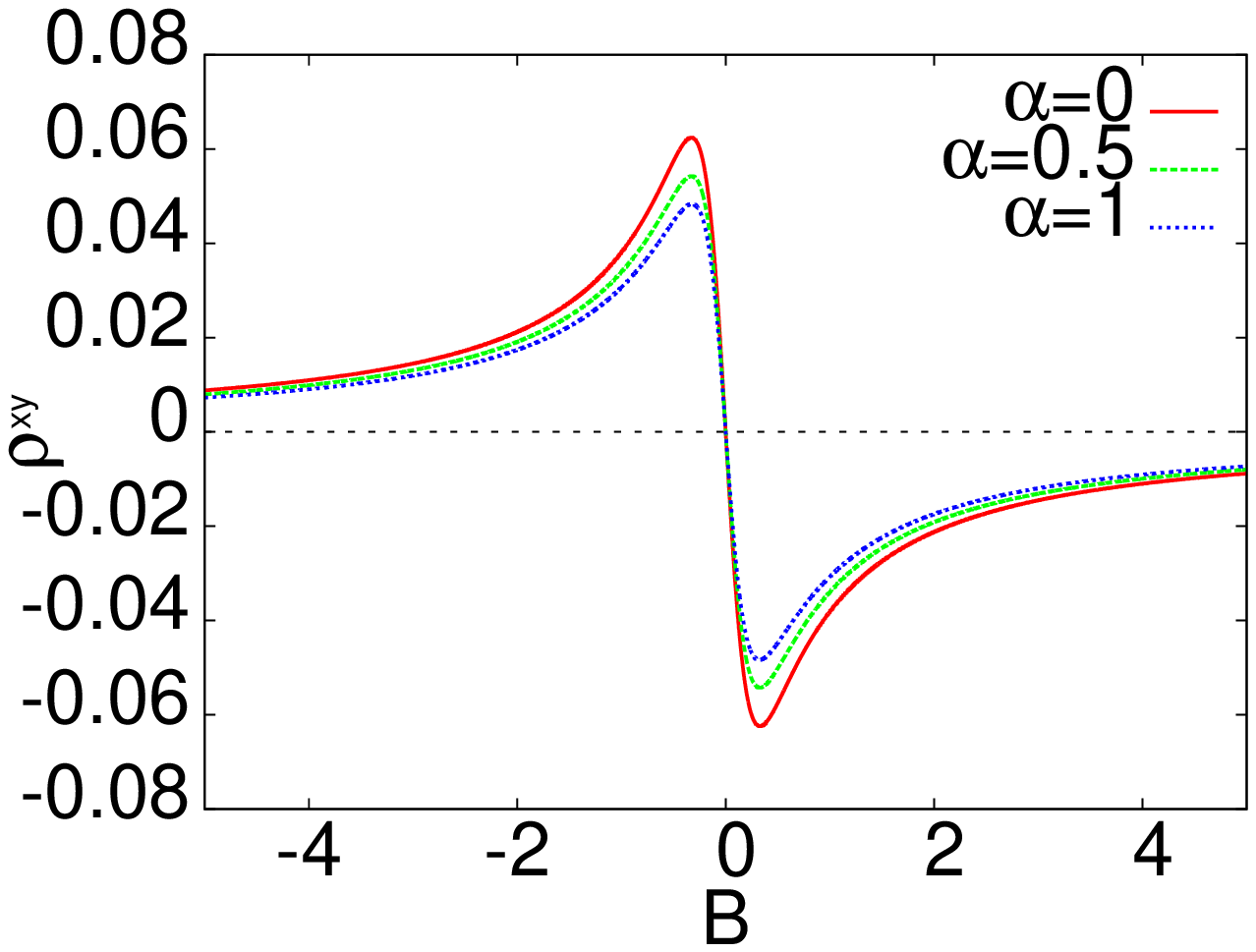}
\caption{ Magnetic field  dependence of the resistivity 
$ \rho^{xx} $ (left panel) and $\rho^{xy} $ (right panel) 
for three values of $\alpha=0,0.5,1$ and other parameters set to  $n=0.1$, $g=0$  and $\mu=1$.}
\label{fig1-8}
\end{figure}

The presented calculations are  valid  for systems which
can be considered as a strongly coupled ones.
According to the earlier discussion, this condition is expected to be valid
relatively close to the Dirac point.
In \cite{liang2017} both Hall and Nernst effects were analyzed in terms 
of anomalous contributions arising solely from the Berry phase.   
Our approach can not quantitatively describe the Berry phase induced contribution to the transport. 
On the contrary, it gives strong coupling 
contributions relevant for systems with high mobility.

It is expected that in the studied materials  the inter-valley scattering might contribute
to the transport. However,
unlike the graphene, even strong spin - orbit coupling does not gap the spectrum in DSM. 
The oscillations of the transport coefficients in high magnetic fields are also envisaged in real systems.  However, this 
effect
we do not take into account in the holographic approach. The same is true for the interference effects conjectured due to
the Berry phases from electron and hole sheets, at Fermi energy. These are the reasons why
our results compare  only qualitatively with real data \cite{liang2017}.

\section{Summary and conclusions}
\label{sum-concl}
{In summary, we have calculated the transport properties of the three dimensional analog of the
graphene, usually called Dirac semi-metal (DSM), by using gauge/gravity duality. Motivated by the previous
approach to the graphene transport we have  generalized the approach by allowing for the mixing
of two currents as expressed {\it via} the term in the action (\ref{sgrav}) proportional to $\alpha$. 
We have also introduced magnetic field ${\bf B=(0,~0,~B_z)}$ directed perpendicularly to the electric fields ${\bf E_F}$
and ${\bf E_B}$ and the temperature gradient $\nabla T$, all applied in the (x,y) plane. 
The obtained results generalize the known Drude-Boltzmann-like formula to the strong coupling limit. 
This shows up as an additional term $(\mu B)^2~(\tQ_{(F)} \mu r^2_h)^2$
 appearing in the denominator of the Drude like formula for conductance $\sigma(B)$, where
 $\mu^2=1/(12~\beta^2~r_h^2)$ plays on the holographic side the role of impurity limited mobility of charges in the
DSM under consideration. }

\acknowledgments
MR was partially supported by the grant no. DEC-2014/15/B/ST2/00089 of the National Science Center 
and KIW by the grant DEC-2014/13/B/ST3/04451.



\end{document}